\documentclass[pdflatex,sn-mathphys]{sn-jnl}

\usepackage{mathtools}
\usepackage{cleveref}
\usepackage[symbol]{footmisc}

\def\vec   #1{\mbox{\boldmath $#1$}{}}
\def\ten   #1{\mbox{\boldmath $#1$}{}}

\jyear{2022}
\begin{document}

\title[PINNs for cardiac fibers]{Physics-informed neural networks to learn cardiac fiber orientation from multiple electroanatomical maps}

\author[1,4]{\fnm{Carlos} \sur{Ruiz Herrera}$^{\dagger,}$}\email{caruiz2@uc.cl}
\author[2,6]{\fnm{Thomas} \sur{Grandits}$^{\dagger,}$}\email{thomas.grandits@uni-graz.at}
\author[6]{\fnm{Gernot} \sur{Plank}}\email{gernot.plank@medunigraz.at}
\author[3]{\fnm{Paris} \sur{Perdikaris}}\email{pgp@seas.upenn.edu}
\author*[1,4]{\fnm{Francisco} \sur{Sahli Costabal}}\email{fsc@ing.puc.cl}
\author[5]{\fnm{Simone} \sur{Pezzuto}}\email{simone.pezzuto@usi.ch}

\affil[1]{\orgdiv{Department of Mechanical and Metallurgical Engineering},
\orgdiv{School of Engineering},
\orgname{Pontificia Universidad Cat\'olica de Chile},
\orgaddress{\city{Santiago}, \country{Chile}}}

\affil[2]{\orgdiv{Institute of Mathematics and Scientific Computing},
\orgname{University of Graz},
\orgaddress{\city{Graz}, \country{Austria}}}

\affil[3]{\orgdiv{Department of Mechanical Engineering and Applied Mechanics},
\orgname{University of Pennsylvania},
\orgaddress{\city{Philadelphia}, \country{USA}}}

\affil[4]{Institute for Biological and Medical Engineering, Schools of Engineering, Medicine and Biological Sciences, \orgname{Pontificia Universidad Cat\'olica de Chile},
\orgaddress{\city{Santiago}, \country{Chile}}}

\affil[5]{\orgdiv{Center for Computational Medicine in Cardiology},
\orgdiv{Euler Institute},
\orgname{Universit\`a della Svizzera italiana},
\orgaddress{\street{via Buffi 13}, \city{Lugano}, \postcode{6900}, \country{Switzerland}}}

\affil[6]{\orgdiv{Gottfried Schatz Research Center - Division of Biophysics},
\orgname{Medical University of Graz},
\orgaddress{\city{Graz}, \country{Austria}}}


\abstract{We propose FiberNet, a method to estimate \emph{in-vivo} the cardiac fiber
architecture of the human atria from multiple catheter recordings
of the electrical activation. Cardiac fibers play a central role
in the electro-mechanical function of the heart, yet they are
difficult to determine \emph{in-vivo}, and hence rarely truly patient-specific
in existing cardiac models.
FiberNet learns the fiber arrangement by solving
an inverse problem with physics-informed neural networks. The inverse problem amounts to identifying
the conduction velocity tensor of a cardiac propagation model
from a set of sparse activation maps. The use of multiple maps
enables the simultaneous identification of all the components
of the conduction velocity tensor, including the local fiber angle.
We extensively test FiberNet on synthetic 2-D and 3-D examples, diffusion tensor fibers, and a patient-specific case. We show that 3 maps are sufficient to accurately capture the fibers, also in the
presence of noise. With fewer maps, the role of regularization becomes
prominent. Moreover, we show that the fitted model can robustly
reproduce unseen activation maps. We envision that FiberNet will help the creation of patient-specific models for personalized medicine.
The full code is available at \url{http://github.com/fsahli/FiberNet}.}

\keywords{Cardiac Fibers, Physics-Informed Neural Networks,
Cardiac Electrophysiology, Anisotropic conduction velocity,
Eikonal Equation, Deep learning}

\renewcommand*{\thefootnote}{\fnsymbol{footnote}}
\footnotetext[2]{These authors contributed equally to this work}
\renewcommand*{\thefootnote}{\arabic{footnote}}
\setcounter{footnote}{0}

\maketitle


\section{Introduction}\label{sec:intro}


In recent years, the future vision of precision cardiology through digital twinning has gained traction~\cite{corral_acero_digital_2020,peirlinck2021precision}. In such scenarios, a digital replica of a patient's heart is generated from a variety of measurements in the clinic. Such replica can be used in a multitude of ways, from predicting intervention outcomes to building cohorts for augmented drug trials. However, building such cardiac digital twins faces several difficulties, such as dealing with noisy measurements, or the problem of having multiple possible parameterizations explaining the encountered measurements, mathematically known as ill-posedness \cite{o1986statistical}.

The fiber distribution in the heart is a key determinant of cardiac function and, as such, it has a prominent role in digital twinning~\cite{ho2009importance}. The electrical stimulus activating the myocardium travels at different speeds, depending on whether the propagation occurs along or across the fibers~\cite{clerc1976directional}. More precisely, the electrical conduction within the myocardium is \emph{anisotropic}, because the electrical propagation is faster in the direction of the fibers.  There are multiple determinants of anisotropic conduction,
and some may uncover a pathological condition~\cite{kotadia2020anisotropic}. Therefore, an imprecise knowledge of the anisotropic conduction in digital twinning may yield wrong predictions or even non-physiological behaviors.

In patient-specific models, cardiac fibers are generally arranged following some rules, dictated by prior histological knowledge on their distribution in the heart~\cite{streeter1969fiber,ho2009importance}. These algorithms are well-established for the ventricles~\cite{Bayer2012}. More recently, some rule-based methods and atlas-based for the atria have also been proposed~\cite{gonzales2013three,Wachter2015Fibers,Roney2019UAC,roney2021constructing,Piersanti2021Fibers}. In all cases, the fiber field is anatomically-tailored, but technically speaking not yet patient-specific. {Diffusion-Tensor Magnetic Resonance (DT-MR) imaging of the heart, the gold-standard tool for sub-millimeter imaging of cardiac fibers~\cite{lombaert2012human,teh2016resolving,pashakhanloo2016myofiber}, is unfortunately not feasible \emph{in-vivo} for the atrial wall.}

A clinically-viable way to infer the local fiber direction is based on the conductivity properties of the tissue. Clinically, the conduction velocity (CV) can be indirectly determined with an electroanatomical mapping system~\cite{Cantwell2015,Coveney2020GP}, a minimally-invasive, catheter-based tool to record the local activation of the endocardial wall. Electroanatomical maps are common in clinical electrophysiology, because they are routinely acquired before ablation therapy, and are relatively easy to export in text format for further inspection. By comparing the local CV obtained from multiple maps, it is therefore possible to derive the direction of fastest conduction, that is the fiber direction~\cite{roney2019technique}.

In this work, we similarly aim at determining the fibers from multiple maps.  Rather than working locally, we solve an inverse problem where the conductivity tensor can be deduced from the activation map through a propagation model, namely the eikonal equation~\cite{franzone1990wavefront,Grandits2021PIEMAP}. This formulation has a number of advantages. First, as the fiber field is recovered, we simultaneously fit a predictive model that, in principle, can faithfully reproduce the observed activations. Second, the model can extrapolate, in a physiological manner, the activation map in regions where data is scarce {or absent~\cite{Lubrecht2021PIEMAP}}. Third, the inverse problem can be easily informed by prior histological knowledge on the fiber distribution through a regularization term or, more generally, a Bayesian approach, {for instance by using rule-based fiber fields~\cite{Piersanti2021Fibers}. Therefore, the fiber field could be extrapolated as well in a consistent and physiological manner.}


Physics Informed Neural Networks (PINNs) are a recently developed variant of machine learning-based methods to efficiently solve inverse problems that are governed by partial differential equations~\cite{raissi_pinn_2019}. They have been shown to accurately model complex physical problems with a small number of known data points. Unlike regular neural networks which typically require vast amounts of labeled data to make accurate predictions, PINNs can learn much quicker to simulate this kind of system due to the incorporation of physical laws -- represented as systems of partial differential equations -- into their loss functions. Moreover, PINNs are a genuinely mesh-free method, overcoming the issue of generating meshes of complex domains like the heart. {In fact, the model and data are penalized at collocation and data points, respectively, while neural networks representing the quantities of interests can be evaluated at any spatial location. Thus, no topological information from the geometrical model is required.}

\begin{figure}[tb]
    \centering
    \includegraphics[width=\textwidth,trim=0 60 0 60,clip]{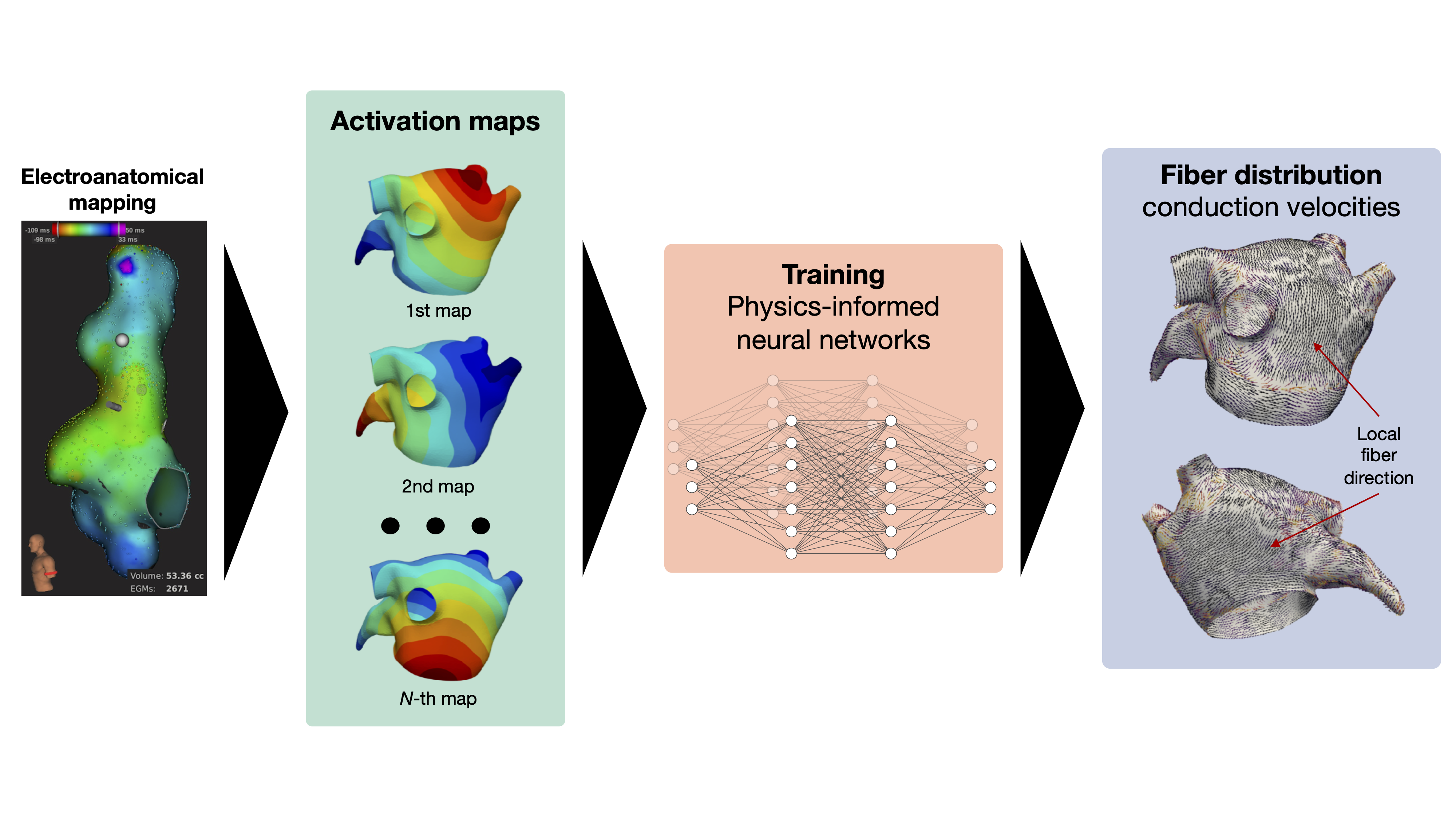}
    \caption{FiberNet translates a set of electroanatomical maps into a continuous estimate of the fiber field and conduction velocity. These estimates can be used for simulating cardiac activation in a predictive model based on the eikonal equation. Internally, FiberNet uses physics-informed neural networks to constrain the parameter space during the training phase.}
    \label{fig:intro}
\end{figure}

In this work, we propose FiberNet, a PINN-based method to solve the inverse problem of identifying fibers in the heart from electroanatomical maps (Fig.~\ref{fig:intro}). This is achieved by simultaneously fitting multiple neural networks to multiple electroanatomical maps while using a common network that predicts the conduction velocity tensor at different locations.
{In this sense, FiberNet extends our previous work on the same problem~\cite{Grandits2021PIEMAP,Lubrecht2021PIEMAP,grandits_learning_2021}, which only used a single activation map and was based on a different representation of the conductivity tensor.}
We extensively validate FiberNet with several numerical experiments and real data. The first set of experiments consists of a completely synthetic example with a 2-D planar geometry and a 3-D atrial geometry, where we compare the single- and multi-map approaches. Secondly, we apply FiberNet to an atlas of diffusion tensor image fiber fields of human atria~\cite{roneyAtlas}. {Here, the ground-truth activation times are generated with the Fast Iterative Method (FIM) simulation for triangular surfaces~\cite{fu_fast_2011,grandits_fast_2021}.} Moreover, we validate the predictive capabilities of FiberNet by first fitting the fibers on training maps, and then generating an unseen map and attempting to predict it using the learned fibers. Finally, we apply FiberNet to clinically-obtained EAMs of the left atrium, thus proving the feasibility of the approach in the clinical setting.


This paper is structured as follows. In Section~\ref{sec:related} we review some of the existing methods to estimate the conduction velocity from sparse activation recordings, a highly relevant topic for this work. FiberNet is introduced in Section~\ref{sec:methods}, along with the PINN framework. The identifiability of the fibers from single and multiple maps and the role of the regularization are discussed in Section~\ref{sec:identify}. The last part of our work is devoted to numerical validation and applications. In Section~\ref{sec:assessment} we verify the robustness to noise of FiberNet in a synthetic 2-D example and on a 3-D atrial geometry.  An extensive benchmark with a publicly available atrial fiber atlas~\cite{roney2021constructing,roneyAtlas} is given in Section~\ref{sec:dtmri}. Finally, in Section~\ref{sec:patspec} we apply FiberNet to a clinical data set. A discussion and outlook are provided in Section~\ref{sec:conclusion}.


\section{Related works}
\label{sec:related}

The problem of identifying cardiac fibers from electrical data of atrial activation has been considered only very recently, see e.g.~\cite{roney2019technique,Grandits2021PIEMAP,Lubrecht2021PIEMAP,grandits_learning_2021}. More broadly, however, the topic is related to the identification of the local conduction velocity (CV) of the tissue, for which several methods have been proposed thus far. In this section, we briefly review the most common approaches. We group them into two categories: data-driven and physics-driven methods. In the former, the CV is directly estimated from the data through either geometrical or statistical arguments; in the latter, the conduction parameters of a cardiac {model} are fitted to the data, hence the CV is a byproduct of the forward model.

\subsection{Data-driven methods}
\label{sec:datadriven}

Trivially, the CV is the ratio between the distance traveled by an activation front in a given amount of time. The activation time can be detected from local electrograms obtained with contact mapping systems. The location of the electrodes on the catheter is also tracked with good accuracy. Therefore, with sufficient coverage of the atrial surface, it is possible to estimate the local CV in a purely data-driven manner, with no prior assumption on the underpinning physiology.

A first class of methods estimates the local CV $\theta(\vec{x})$ at some given location $\vec{x}$ on the basis of the temporal differences in activation between $\vec{x}$ and its neighbors~\cite{Cantwell2015,Verma2018CV,vanSchie2021CV,Pagani2021AF,Good2021CV}. These methods are easy to implement and very fast to execute. Here, the main challenge is dealing with noise and inconsistency in the data. Both the location and activation time are subject to noise, hence their incremental ratio, the CV, is greatly susceptible to large variations due to uncertainty. In this scenario, least-squares approaches~\cite{vanSchie2021CV,Pagani2021AF} may perform better than purely geometric methods~\cite{Cantwell2015,Verma2018CV}. Nonetheless, a careful pre-processing of the input data is always necessary, in order to avoid unrealistically fast or low CVs~\cite{Cantwell2015,Nothstein2021CVAR}. In some cases, very high CVs may actually be due to physiological phenomena, such as breakthroughs and front collisions where, however, the CV is not well-defined.

Rather than estimating the CV from point-wise measurements, other authors suggested to first interpolate or recover a smooth activation map from the data, say $\tilde{\phi}(\vec{x})$, and then compute the conduction velocity from its definition, that is
\begin{equation}\label{eq:CV}
\theta(\vec{x})=\|\nabla\tilde{\phi}\|^{-1}.
\end{equation}
This has been recently done with Gaussian Process Regression (GPR) on manifolds~\cite{Coveney2020GP}. In GPR, the properties of the kernel (smoothness, correlation length) can be optimized to capture the physics of the problem and reduce generalization error. Furthermore, GPR is probabilistic by nature and enables uncertainty quantification in CV estimates and active learning~\cite{sahli2020physics}.

Since cardiac conduction is anisotropic~\cite{kotadia2020anisotropic}, the value of $\theta(\vec{x})$ also depends on the propagation direction $\vec{p}$. In all previous methods, therefore, the estimated CV depends on the activation map, since $\vec{p}=\nabla\tilde{\phi}(\vec{x})/\|\nabla\tilde{\phi}(\vec{x})\|$.  With the use of multiple activation maps, or by accounting for prior knowledge on the fiber distribution, it is however possible to simultaneously estimate the longitudinal and transverse CV, as recently proposed in~\cite{roney2019technique}. In this case, it is tacitly assumed that the CV in the myocardium obeys a Riemannian metric, that is $\theta(\vec{x},\vec{p}) = \sqrt{\ten{D}(\vec{x})\vec{p}\cdot\vec{p}}$, for some symmetric, positive definite tensor field $\ten{D}(\vec{x})$. In the 2-dimensional case, as for the atrial surface, the tensor $\ten{D}(\vec{x})$ is determined by three independent parameters, e.g., the fiber angle and the CV along and across the fibers. Therefore, at least 3 independent activation maps are required to uniquely identify $\ten{D}$.

Interestingly, from Eq.~\eqref{eq:CV} with $\theta(\vec{x},\vec{p})$ as above, we recover the \emph{anisotropic eikonal equation}~\cite{franzone1990wavefront}, see also Eq.~\eqref{eq:Eik} in Sec.~\ref{sec:learning}.  Therefore, $\ten{D}(\vec{x})$ may be recovered by imposing Eq.~\eqref{eq:CV} for multiple activation maps $\tilde{\phi}_i(\vec{x})$, $i=1,\ldots,M$ at each point $\vec{x}$ of the surface~\cite{roney2019technique}. The corresponding algebraic system for $M=3$, however, may have no unique solution. For instance, this may happen when $\nabla\tilde{\phi}_i(\vec{x})$ and $\nabla\tilde{\phi}_j(\vec{x})$, for $i\neq j$, are parallel at some $\vec{x}$.  Similarly, it may happen that the fitted $\ten{D}(\vec{x})$ is not positive-definite, hence there is no physiological solution.  An alternative approach, valid for an arbitrary number of maps and always ensuring at least one solution, is based on minimizing the residual
\begin{equation}\label{eq:minres}
\sum_{i=1}^M \Bigl( \sqrt{\ten{D}(\vec{x})\nabla\tilde{\phi}_i(\vec{x})\cdot\nabla\tilde{\phi}_i(\vec{x})} - 1 \Bigr)^2,
\end{equation}
on the space of parameters defining $\ten{D}$. It is worth noting that finding $\ten{D}$ can be understood as a fitting-an-ellipse-to-points problem, for which more robust algorithms are available~\cite{gander1994least}.

Finally, it is worth mentioning that there exist methods for estimating the CV directly from the temporal dynamics of the electrograms, without the need for the activation map~\cite{masse2016resolving, Gaeta2021DELTA}.

\subsection{Physics-driven methods}

Data-driven methods may suffer when data is scarce or unevenly distributed, as it often happens with electroanatomical maps. In absence of data, prior physiological knowledge may, however, be enforced to still recover a plausible CV. Physics-driven (or physiology-driven) approaches follow this path. For cardiac electrophysiology, several propagation models can be used to constrain the CV to the activation map. These models can be based on the eikonal equation~\cite{sahli2020physics,Grandits2021PIEMAP,Lubrecht2021PIEMAP,grandits_learning_2021}, reaction-diffusion systems~\cite{yang2015estimation,Barone2020Conductivity,Irakoze2021Dist}, or a multi-fidelity combination thereof~\cite{chegini2021efficient}. The typical workflow aims at minimizing the observational residual between simulated and recorded activation, by optimizing the (distributed) conductivity parameters of the model. Once the optimal parameters of the model have been obtained, the local CV is trivially computed.

Physics-driven approaches are generally more robust than purely data-driven methods, as they allow one to weigh data fidelity against model fidelity, through a regularization term. More importantly, physics-driven methods potentially provide a predictive model of cardiac electrophysiology, thus they can be employed in personalized therapeutic approaches~\cite{Arevalo2016}. However, they have a significantly higher computational footprint, both in terms of memory and time. Moreover, some other parameters, such as the early activation sites, may potentially influence the CV reconstruction~\cite{Grandits2021PIEMAP}.  In this respect, a good trade-off between purely data-driven and physics-driven methods consists in accounting for the physics only weakly~\cite{raissi_pinn_2019,sahli2020physics,grandits_learning_2021}, for instance through a penalization term in the loss function such as Eq.~\eqref{eq:minres}, rather than enforcing the model point-wise. This observation further motivates the method presented below.


\section{Methods}
\label{sec:methods}


\subsection{Propagation model}
\label{sec:model}


Let $\mathcal{S}\subset\mathbb{R}^3$ be a smooth orientable surface representing, for instance, the left atrial endocardium.  We model cardiac activation with the anisotropic eikonal equation~\cite{franzone1990wavefront}. We do not consider diffusion or curvature correction terms. The eikonal equation models the arrival times $\phi(\vec{x})$ resulting from the spread of an electric activation wavefront within the myocardium, propagating with direction-dependent CV. The equation reads as follows
\begin{equation}\label{eq:Eik}
\sqrt{\ten{D}(\vec{x})\nabla\phi(\vec{x})\cdot\nabla\phi(\vec{x})} = 1,
\end{equation}
where $\ten{D}(\vec{x})\in\mathbb{R}^{3\times 3}$ is a symmetric, positive-definite tensor field representing the conductivity. Specifically,
\begin{equation}
\theta(\vec{x},\vec{p}) = \sqrt{\ten{D}(\vec{x})\vec{p}\cdot\vec{p}}
\label{eq:thetaCV}
\end{equation}
is the conduction velocity in direction $\vec{p}$.

We define the local fiber direction as the direction of fastest propagation and we denote it by $\vec{l}(\vec{x})\in\mathbb{R}^3$. Since the propagation is constrained on the atrial surface, $\vec{l}(\vec{x})$ is orthogonal to $\vec{n}(\vec{x})$, the normal direction of $\mathcal{S}$. A generic way to represent $\vec{D}$ is therefore as follows:
\begin{equation}
\ten{D}(\vec{x}) = v_l^2(\vec{x})\,\vec{l}(\vec{x})\otimes \vec{l}(\vec{x})
+ v_t^2(\vec{x})\,\vec{t}(\vec{x})\otimes \vec{t}(\vec{x}),
\label{eq:cgcon}
\end{equation}
where $\vec{t}(\vec{x})\in\mathbb{R}^3$ is the transverse vector field, orthogonal to both $\vec{l}$ and $\vec{n}$, while $v_l$ and $v_t$ are the conduction velocities respectively along $\vec{l}$ and $\vec{t}$. We note that the velocity in the normal direction is zero.

Given an orthonormal basis of the tangent space of $\mathcal{S}$ at a location $\vec{x}$, denoted by $\{ \vec{v}_1(\vec{x}), \vec{v}_2(\vec{x}) \}$, we formulate the fiber and transverse directions as
\begin{alignat}{4}
\vec{l}(\vec{x}) &={}& a\,& \vec{v}_1(\vec{x})\otimes \vec{v}_1(\vec{x}) &{}+{}& &\sqrt{1-a^2}\,& \vec{v}_2(\vec{x})\otimes \vec{v}_2(\vec{x}), \\
\vec{t}(\vec{x}) &={}& - \sqrt{1-a^2}\,& \vec{v}_1(\vec{x})\otimes \vec{v}_1(\vec{x}) &{}+{}& &a\,& \vec{v}_2(\vec{x})\otimes \vec{v}_2(\vec{x}),
\label{eq:tt}
\end{alignat}
where $a\in[-1,1]$ is the cosine of the angle between the fiber direction and $\vec{v}_1(\vec{x})$.
While any orthonormal basis on the tangent space would be feasible (even piecewise-constant), later regularization techniques greatly benefit from the smoothness of the basis. We therefore employed the vector heat method~\cite{sharp2019vector}, implementing parallel transport of vectors on manifolds, to propagate a first initial vector across the entire atrial surface, thus to obtain $\vec{v}_1(\vec{x})$. The second vector field $\vec{v}_2(\vec{x})$ is obtained by orthogonalization. See also Appendix~\ref{apx:smooth} and Figure~\ref{fig:smoothbasis}.

In summary, the conductivity tensor is defined by 3 scalar fields,
\begin{itemize}
    \item $a(\vec{x}) \in [-1,1]$, the cosine of the fiber angle in the $\{ \vec{v}_1, \vec{v}_2 \}$ basis,
    \item $e_1(\vec{x}) \coloneqq v_l^2(\vec{x}) \ge 0$, the square of the longitudinal velocity, and
    \item $e_2(\vec{x}) \coloneqq v_t^2(\vec{x}) \ge 0$, the square of the transversal velocity.
\end{itemize}

\subsection{Learning fibers from multiple maps}
\label{sec:learning}


We consider the problem of simultaneously identifying $a(\vec{x})$, $e_1(\vec{x})$ and $e_2(\vec{x})$ from a set of $M\ge 1$ electroanatomical maps. For convenience, we define
\[
\vec{d}(\vec{x}) = \bigl[ a(\vec{x}), e_1(\vec{x}), e_2(\vec{x}) \bigr],
\]
so that the conductivity tensor $\ten{D}(\vec{x})$, through Eq.~\eqref{eq:cgcon}-\eqref{eq:tt}, is a function of $\vec{d}(\vec{x})$, that is
\[
\ten{D}(\vec{x}) = \tilde{\ten{D}}\bigl( \vec{d}(\vec{x}) \bigr).
\]
We solve the identification problem with the PINN framework, extending our previous work~\cite{grandits_learning_2021}. {Here, we approximate \emph{each} activation map $\phi_m(\vec{x})$, $m=1,\ldots,M$, and the conductivity parameters vector $\vec{d}(\vec{x})$ with a total number of $M+1$ artificial neural networks}, 
\begin{align*}
\phi_m(\vec{x}) &\approx \hat{\phi}_m(\vec{x}) \coloneqq \mathrm{NN}(\vec{x};\vec{\theta}_{\phi_m}), \ \ m=1,\ldots,M\\
\vec{d}(\vec{x}) &\approx \hat{\vec{d}}(\vec{x}) \coloneqq \mathrm{NN}(\vec{x};\vec{\theta}_D),
\end{align*}
where $\vec{\theta}_{\phi_m}$ and $\vec{\theta}_D$ denote the trainable parameters of the networks. The approximated conductivity tensor trivially follows, $\hat{\ten{D}}(\vec{x}) = \tilde{\ten{D}}\bigl( \hat{\vec{d}}(\vec{x}) \bigr)$. {The output layer of $\hat{\phi}_m(\vec{x})$ is the sigmoid function, hence it ranges in $[0,1]$, whereas for $\hat{\vec{d}}(\vec{x})$ we consider a sigmoid scaled by a factor $C > 0$ for $e_1$ and $e_2$, and a hyperbolic tangent activation function for the angle $a$.}

\begin{figure}[htb]
    \centering
    \includegraphics[width = .9 \textwidth]{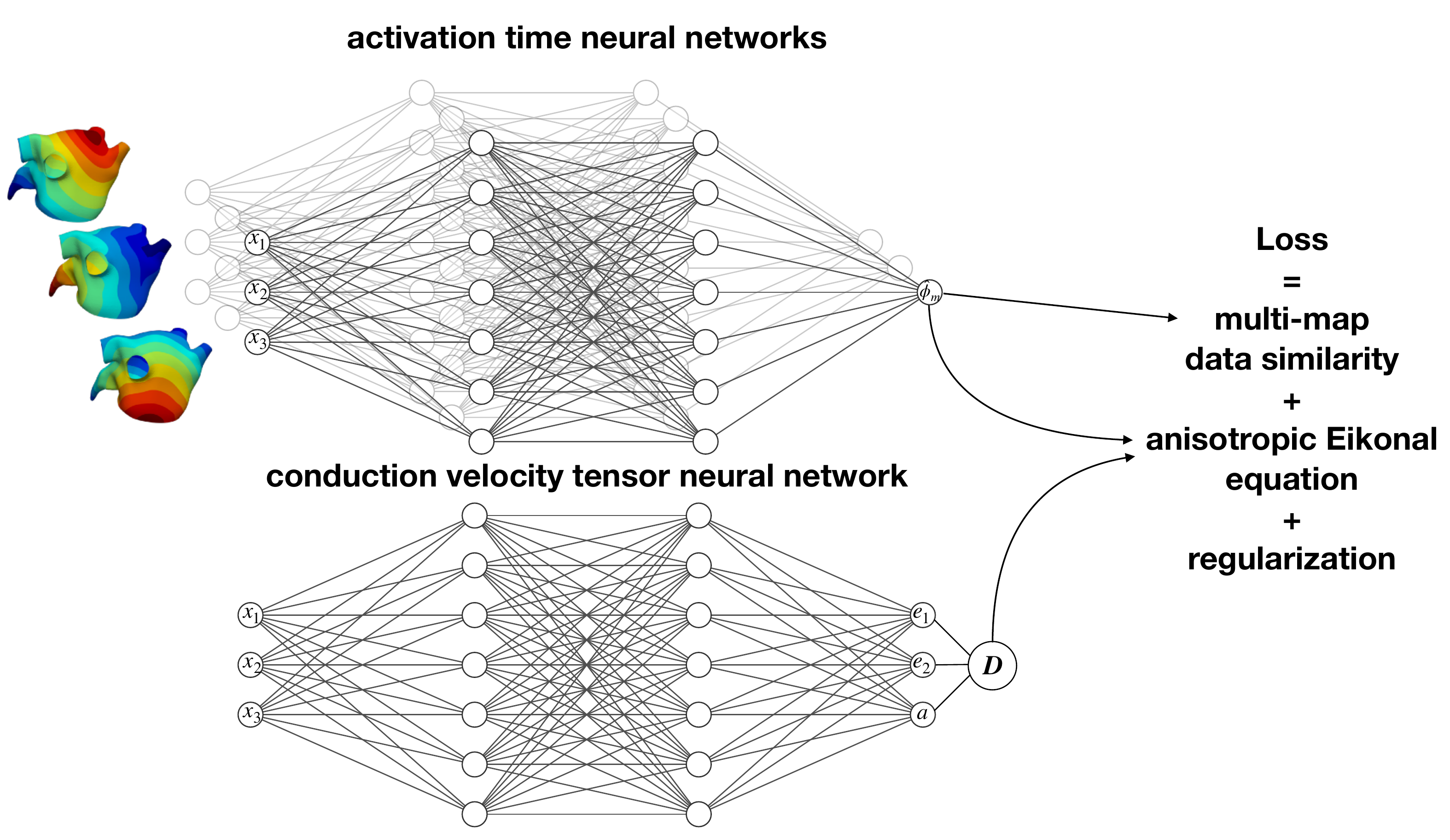}
    \caption{Schematic representation of the physics-informed neural networks employed in this work. We use multiple neural networks to represent each of the activation times received for training. We use one neural network to represent the conduction velocity tensor. From here, we can compute the fiber orientation. We train all the networks simultaneously to satisfy the data from the activation time maps and the Eikonal equation, which links activation times to conduction velocities.}
    \label{fig:net}
\end{figure}

To train our model, we take into consideration the data from the maps, the residual from the eikonal equation, and regularization terms. In Figure~\ref{fig:net} we present a schematic of our approach. We build a loss function that includes all these components, as follows:
{%
\begin{equation}
\label{eq:loss}
\mathcal{L}(\vec{\theta}_\phi, \vec{\theta}_{D}) \coloneqq
\mathcal{L}_\mathrm{data} + \alpha_m \mathcal{L}_\mathrm{eiko} + \alpha_e \mathcal{L}_\mathrm{cv} + \alpha_a \mathcal{L}_\mathrm{ang},
\end{equation}
where $\alpha_m$, $\alpha_e$ and $\alpha_a$ are hyper-parameters. The hyper-parameter $\alpha_m$ controls the relevance of eikonal equation, while the hyper-parameters $\alpha_e$ and $\alpha_a$ control the amount of regularization applied to the conduction velocities and angles, respectively. We also collect all the neural network weights $\{ \vec{\theta}_{\phi_m} \}_{m=1}^N$ in a single vector $\vec{\theta}_\phi$. 

In detail, we consider a set of data points, denoted by $\vec{x}_{m,i}\in\mathcal{S}$, where $m=1,\ldots,M$ indicates the map number, and $i=1,\ldots,N_m$ the point number. For each point, we consider a recorded activation time, denoted by $\phi_m(\vec{x}_{m,i})$. Each activation map is also divided by a factor $T_\mathrm{max}$, so that the data ranges in $[0,1]$. Furthermore, we have a set of collocation points $\vec{y}_j \in \mathcal{S}$, $j=1,\ldots,N_C$, where the eikonal model is enforced. The terms in the loss function are as follows,
\begin{align*}
\mathcal{L}_\mathrm{data} &= \frac{1}{M} \sum_{m=1}^M \frac{1}{N_m} \sum_{i=1}^{N_m} \left(\hat{\phi}(\vec{x}_{m,i}) - \phi(\vec{x}_{m,i})\right)^2, \\
\mathcal{L}_\mathrm{eiko} &= \frac{1}{M N_C} \sum_{m=1}^M\sum_{j=1}^{N_C} \left(T_\mathrm{max} \sqrt{\hat{\ten{D}}(\vec{y}_j) \nabla\hat{\phi}_m(\vec{y}_j)\cdot \nabla\hat{\phi}_m(\vec{y}_j)} - 1 \right)^2, \\
\mathcal{L}_\mathrm{cv} &= \frac{1}{N_C} \sum_{j=1}^{N_C} \Bigl( H_{\delta_e}\bigl(\nabla e_1(\vec{y}_j))\bigr) + 
H_{\delta_e}\bigl(\nabla e_2(\vec{y}_j))\bigr) \Bigr), \\
\mathcal{L}_\mathrm{ang} &= \frac{1}{N_C} \sum_{j=1}^{N_C} H_{\delta_a}\bigl(\nabla a(\vec{y}_j)\bigr).
\end{align*}
}
We note that the eikonal model is not enforced for all $\vec{x}\in\mathcal{S}$, as generally done in PDE-constrained optimization, but rather on a set of collocation points $\{ \vec{y}_j \}_{j=1}^{N_C}$. In this respect, the method is \emph{mesh-free}, since there is no need for a triangulation of $\mathcal{S}$ to represent the quantities of interest. Moreover, the eikonal equation is never actually solved, an important aspect since we do not have a precise knowledge of the earliest activation sites.

For the regularization terms, we consider the Huber Total Variation function
\begin{equation}
    H_{\delta}(\vec{q}) = 
    \begin{cases}
     \frac{1}{2 \delta} 
     \|\vec{q}\|^2, & \mbox{if $\|\vec{p}\| \le \delta$,} \\
     \|\vec{q}\| - \frac{1}{2} \delta, & \mbox{otherwise}.
    \end{cases}
\label{eq:huber}
\end{equation}
The Huber total variation penalization is used in this problem as an additional restriction for the minimization because it has been shown to favor {piecewise constant solutions~\cite{chambolle_introduction_2016}}. We use a different penalization for the fiber angle and the conduction velocities to be able to have more control over their individual regularity, as shown in Section~\ref{sec:assessment}.




In summary, we train the neural networks with the loss function~\eqref{eq:loss} to find the weights $\vec{\theta}_\phi$ and $\vec{\theta}_D$, with hyper-parameters 
$\alpha_m$, $\alpha_e$, $\alpha_a$, $\delta_e$, $\delta_a$.

\subsection{Identifiability}
\label{sec:identify}

In view of the discussion in Section~\ref{sec:datadriven}, multiple activations are required to reconstruct the fiber field. More precisely, at least 3 maps, along with some conditions, are needed to uniquely determine $\vec{d}(\vec{x})$. We show here that thanks to the proposed regularization mechanisms this is not the case for FiberNet. In fact, we can always obtain an estimate of the fiber field, even from a single map, as done in~\cite{grandits_learning_2021}. However, multiple maps will significantly improve the estimation of the fiber field and the CVs, as we extensively show in Section~\ref{sec:assessment}.

\begin{figure}[tb]
\centering
\includegraphics[width=\textwidth,trim=100 20 80 10,clip]{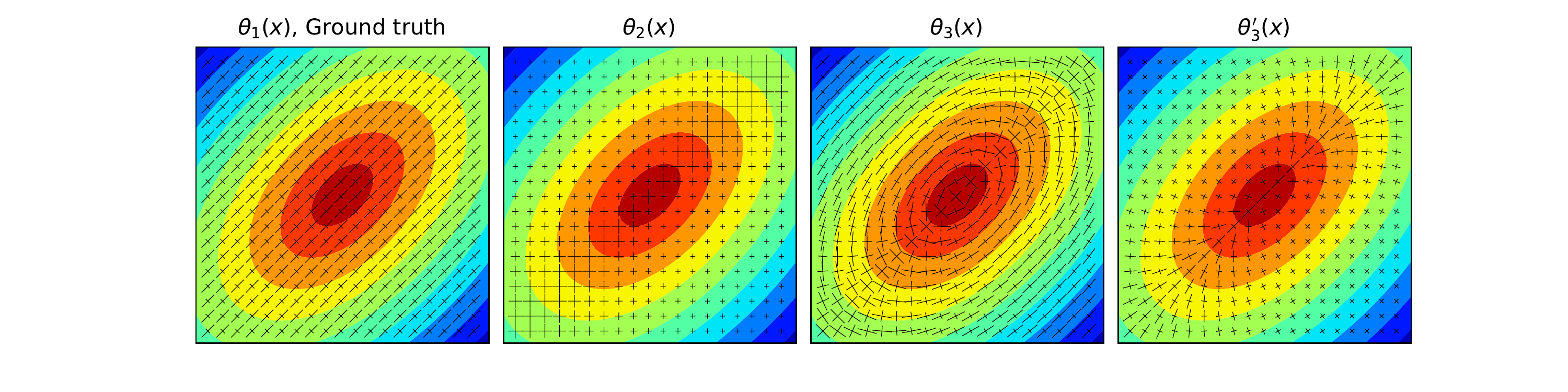}
\caption{Multiple conductivity tensors reproducing the same activation map. The tensor is represented with its eigenvectors rescaled by the eigenvalues. The solution with lowest Huber norm is $\theta_1(\vec{x},\vec{p})$, which also corresponds to the ground truth.}
\label{fig:multisol}
\end{figure}

As an example, let us consider the problem in a 2-D square domain, and suppose that the true conductivity tensor is $\ten{D}\in\mathbb{R}^{2\times 2}$, a symmetric, positive-definite, constant tensor. Here, the fiber direction is the principal eigenvector of $\ten{D}$. If we start an activation at the origin, the true activation map is as follows:
\begin{equation}
\phi(\vec{x}) = \sqrt{\ten{D}^{-1}\vec{x}\cdot \vec{x}}.
\label{eq:sol2Dconst}
\end{equation}
We construct now multiple conductivity tensors yielding the same activation map. We can rewrite the eikonal model as
\[
\theta(\vec{x},\vec{p})\|\nabla\phi(\vec{x})\| = 1, \qquad \vec{p} = \frac{\nabla\phi(\vec{x})}{\|\nabla\phi(\vec{x})\|},
\]
where $\theta(\vec{x},\vec{p})$ is the conduction velocity in the propagation direction $\vec{p}$. Then, as anticipated above, the most trivial choice is:
\[
\theta_1(\vec{x},\vec{p}) = \sqrt{\ten{D}\vec{p}\cdot\vec{p}}.
\]
We can also reproduce Eq.~\eqref{eq:sol2Dconst} with an \emph{isotropic} model but spatially-varying conduction velocity:
\[
\theta_2(\vec{x},\vec{p}) = \sqrt{\frac{\ten{D}^{-1}\vec{x}\cdot \vec{x}}{\ten{D}^{-1}\vec{x}\cdot \ten{D}^{-1}\vec{x}}}.
\]
Even by enforcing a space-varying anisotropic model,
\[
\theta_3(\vec{x},\vec{p}) = \sqrt{\tilde{\ten{D}}(\vec{x})\vec{p}\cdot\vec{p}},
\]
we can still find multiple choices, besides $\tilde{\ten{D}}(\vec{x})=\ten{D}$. Consider the following transversely isotropic tensor:
\[
\tilde{\ten{D}}(\vec{x}) = e_2(\vec{x}) \ten{I} + \bigl( e_1(\vec{x}) - e_2(\vec{x})\bigr)\vec{l}(\vec{x}) \otimes \vec{l}(\vec{x}),
\]
where $\vec{l}(\vec{x})$ is the fiber direction. After substituting it in the previous expression, we have
\[
\theta_3(\vec{x},\vec{p}) = \sqrt{e_2(\vec{x}) + \bigl( e_1(\vec{x}) - e_2(\vec{x})\bigr) \bigl(\vec{l}(\vec{x})\cdot \vec{p}\bigr)^2},
\]
thus, by choosing $\vec{l}$ such that $\vec{l}(\vec{x}) \cdot \vec{p} = 0$ for all $\vec{x}$ and $e_2(\vec{x})=\theta^2_2(\vec{x})$, we obtain a solution for any choice of $e_1(\vec{x}) > e_2(\vec{x})$.  Alternatively, by taking $\vec{l}(\vec{x}) = \vec{p}$ and $e_1(\vec{x})=\theta^2_2(\vec{x})$, we again have an infinite number of solutions by varying $e_2(\vec{x}) < e_1(\vec{x})$.  Therefore, as expected, a single map is not sufficient to fully qualify the tensor $\ten{D}$ from an algebraic point of view. See Figure~\ref{fig:multisol} for a visual example.

Which fiber field is the learning algorithm going to prefer?  Suppose we have a map sampled from Eq.~\eqref{eq:sol2Dconst} in absence of noise. Then, all the above choices for $\theta$ {cancel} the fidelity term \emph{and} the eikonal penalization term in the loss function, because they faithfully reproduce the true activation map. {All these cases are global minima of the loss function~\eqref{eq:loss}, at least when no regularization is present. The neural networks would simply interpolate the data, in a least-squares sense.  In the presence of regularization, however, the landscape of the loss function differs significantly. In fact, $\theta_1$ would likely be favored because, being constant in space, it has the smallest Huber norm amongst all possible choices. However, this choice is mostly dictated by the \emph{prior} assumption encoded by the regularization itself, rather than the data, thus the reconstructed fibers may sensibly differ from the ground truth, especially in the presence of heterogeneity. Therefore, a reliable method for reconstructing the fibers should rely on multiple maps, as proposed in our approach.}



\section{Numerical assessment}
\label{sec:assessment}

We implemented FiberNet using Tensorflow \cite{tensorflow}. For all experiments, we use
4 CPUs of an AMD EPYC 7702 64-Core Processor in parallel for a fixed number of iterations of ADAM~\cite{kingma_adam_2017} with mini-batches and the default hyper-parameters. We ran 4 different sets of experiments: a synthetic 2-dimensional case, {a synthetic 3-dimensional case on a patient-derived geometrical model}, 7 different cases where the fiber orientations {were} measured with {DT-MR imaging}, and one case where 3 different maps were obtained for a patient. 
For each of these experiments, we report both activation time and fiber orientation error. For the activation times, we report the root mean squared error. We quantify the fiber error as: $\frac{1}{N_S}\sum_S \arccos{(\vec{f}\cdot \hat{\vec{f}})}$, where $\vec{f}$ is the ground truth direction of the fibers used to generate data and $\hat{\vec{f}}$ is the predicted fiber direction, which is obtained as the eigenvector associated of with largest eigenvalue of $\hat{\ten{D}}$. The data fidelity error is measured as the root mean squared error (RMSE) of $(\phi-\hat{\phi})$ of all the surface points. For the case of having multiple maps the same RMSE measurement is used but the error is of $(\phi_m-\hat{\phi}_m), \forall m$. {We fix the following hyper-parameters for all these experiments: $\{\alpha_a:10^{-9},\ \delta_e:10^{-3},\ \delta_a:10^{-3}\}$. The remaining hyper-parameters used in each experiment are specified in each section.}

\subsection{Synthetic 2D example}

For the first set of numerical experiments, we use a flat domain defined as $\Omega\coloneqq[-1,1]\times[-1,1]$ discretized with a regularly spaced grid of $35\times 35$ points. We generate a triangular mesh with these points. We create synthetic maps by solving the eikonal equation with a fast iterative method~\cite{grandits_fast_2021}. We select 5 earliest activation sites using a random Latin hypercube design, to avoid selecting points that are close and effectively generate the same map. We set the conduction velocity tensors in the domain with a piece-wise constant function:
\begin{equation}
    \ten{D}(x,y)=
    \begin{cases}
    \begin{bmatrix} 1 & 0 \\ 0 & \frac{1}{2} \end{bmatrix}
    & x + y < 0, \\ 
    \begin{bmatrix} \frac{1}{2} & 0 \\ 0 & 1 \end{bmatrix} & \mbox{otherwise}.\\
    \end{cases}
\label{eq:cond}
\end{equation}

\begin{figure}[tbh]
    \centering
    \includegraphics[width = 0.92\textwidth]{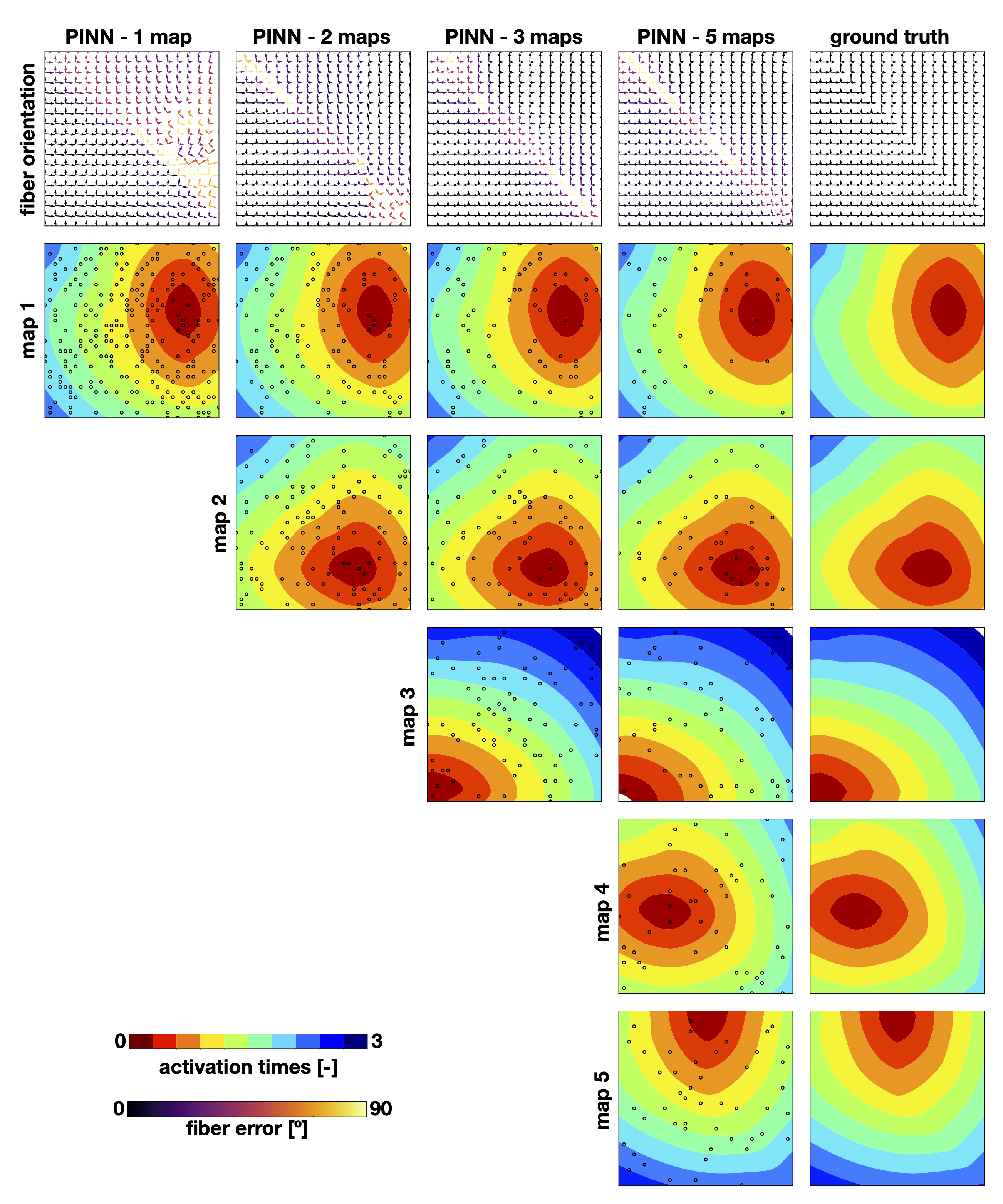}
    \caption{Activation times simulated by the runs with median error for hyper-parameters: $\{\alpha_m:10^{-2},\ \alpha_e:10^{-5},\ \alpha_a:10^{-9},\ \delta_e:10^{-3},\ \delta_a:10^{-3}\}$ for 1, 3 and 5 activation maps and the corresponding estimated conduction velocities (first row). The sampled points used for each run are represented as black circles and the colors are scaled to the 0.0-3.0 range (with white meaning out of range) for the activation time maps and to the range 0-90$^\circ$ for the fiber orientation error in the conduction velocity maps.}
    \label{fig:2Dexample}
\end{figure}

From the solution, we select 245 points, using a Latin hypercube design, as data to train the physics-informed neural networks. We feed the neural networks with either 1, 2, 3, or 5 of the generated maps. We split data points between maps, such that the total amount of points remains constant. The synthetic fiber orientations and the 5 maps used to train the model are shown in Figure~\ref{fig:2Dexample}. We set the hyperparameters to $[\alpha_m : 10^{-2},\ \alpha_a:10^{-9},\ \delta_e:10^{-3},\ \delta_a:10^{-3}]$, while for $\alpha_e$, the regularization of the conduction velocities, we run a sensitivity study between the values  $10^{-9}$ and $10^{-3}]$. We use 5 hidden layers of 10 neurons for each of the networks that predict $\hat{\phi}$ and 5 hidden layers of 5 neurons each to approximate $\hat{\ten{D}}$. Each network is then trained for 3000 Adam mini-batch iterations with batch-size 32. Each of these tests is repeated 5 times using the same set of activation sites and sample points to quantify the variability of the training process.

{In Figure~\ref{fig:2Dexample}, we show the results for a representative case with median error for $\alpha_e = 10^{-5}$}. First, we observe that the quality of the approximation of the activation maps decreases as the number of maps increases since there are fewer data points per map. However, the accuracy in the reconstruction of the maps does not translate into a more precise reconstruction of the fiber field. For instance, when only one map is used, the activation map is near perfectly reconstructed, but we can observe large fiber errors, some of them up to nearly 90$^o$. With 2 maps, the errors are greatly reduced, however, there is a region of large fiber error near the lower right corner. For 3 and 5 maps, the errors are concentrated near the transition in fiber orientation, which is to be expected, as there are not enough data points to clearly define that boundary.

\begin{figure}[ht]
    \centering
    \includegraphics[width = \textwidth]{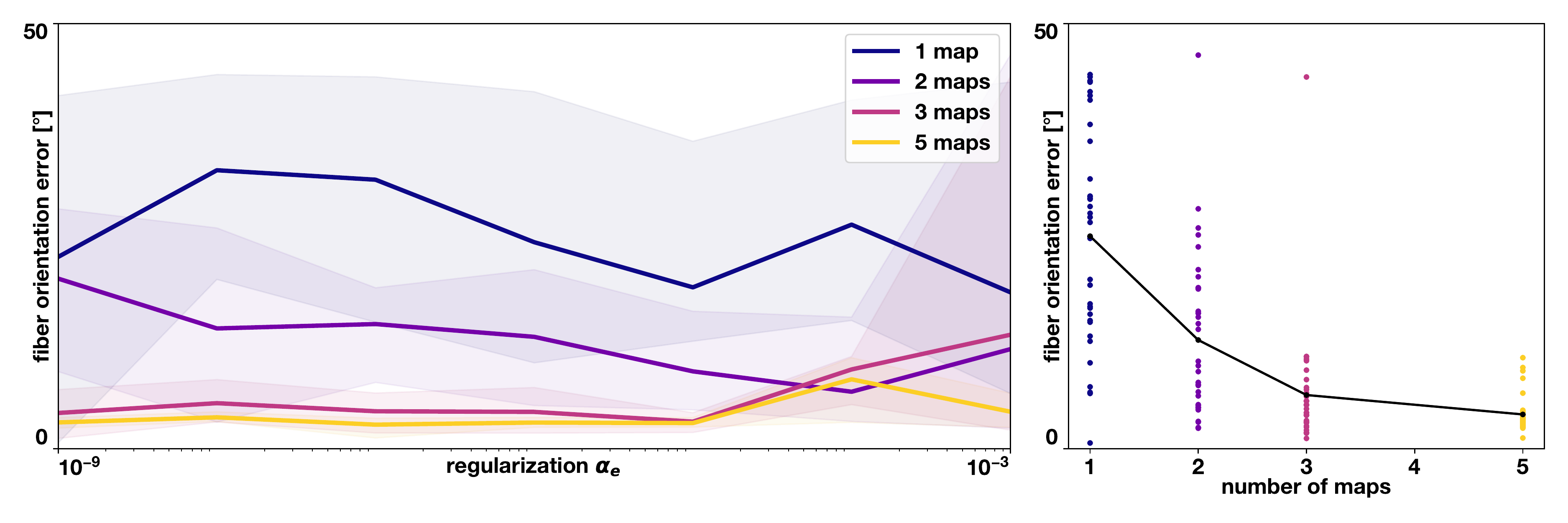}
    \caption{{Left panel: fiber orientation error for $\alpha_m = 10^{-2}$ and $\alpha_{a} = 10^{-9}$. Error values of all runs (shaded areas) and averages for 5 runs (solid lines) in 2D case of PINN with 1, 3 \& 5 maps as inputs using different values of the $\alpha_{e}$ conduction velocity total variation penalization weight. Right panel: fiber orientation errors for all regularization values for different number of maps. The solid line represents the mean.}}
    \label{fig:2D_sens}
\end{figure}

{The fiber orientation errors for varying degrees of $\alpha_e$ are presented in Figure~\ref{fig:2D_sens}}. As shown in the right panel, we observe that for a larger amount of maps, the fiber error is decreased. Nonetheless, the gains from going from 1 to 3 maps are considerably bigger than going from 3 to 5 maps. We also see that the results tend to be more robust to the level of regularization when we feed the model with 3 or 5 maps. The results for 1 and 2 maps show great variability, {which reflects} the ill-posedness of this inverse problem when less than 3 maps are available.

\subsection{Atrial geometry with rule-based fibers}

\begin{figure}[ht]
    \centering
    \includegraphics[width = \textwidth]{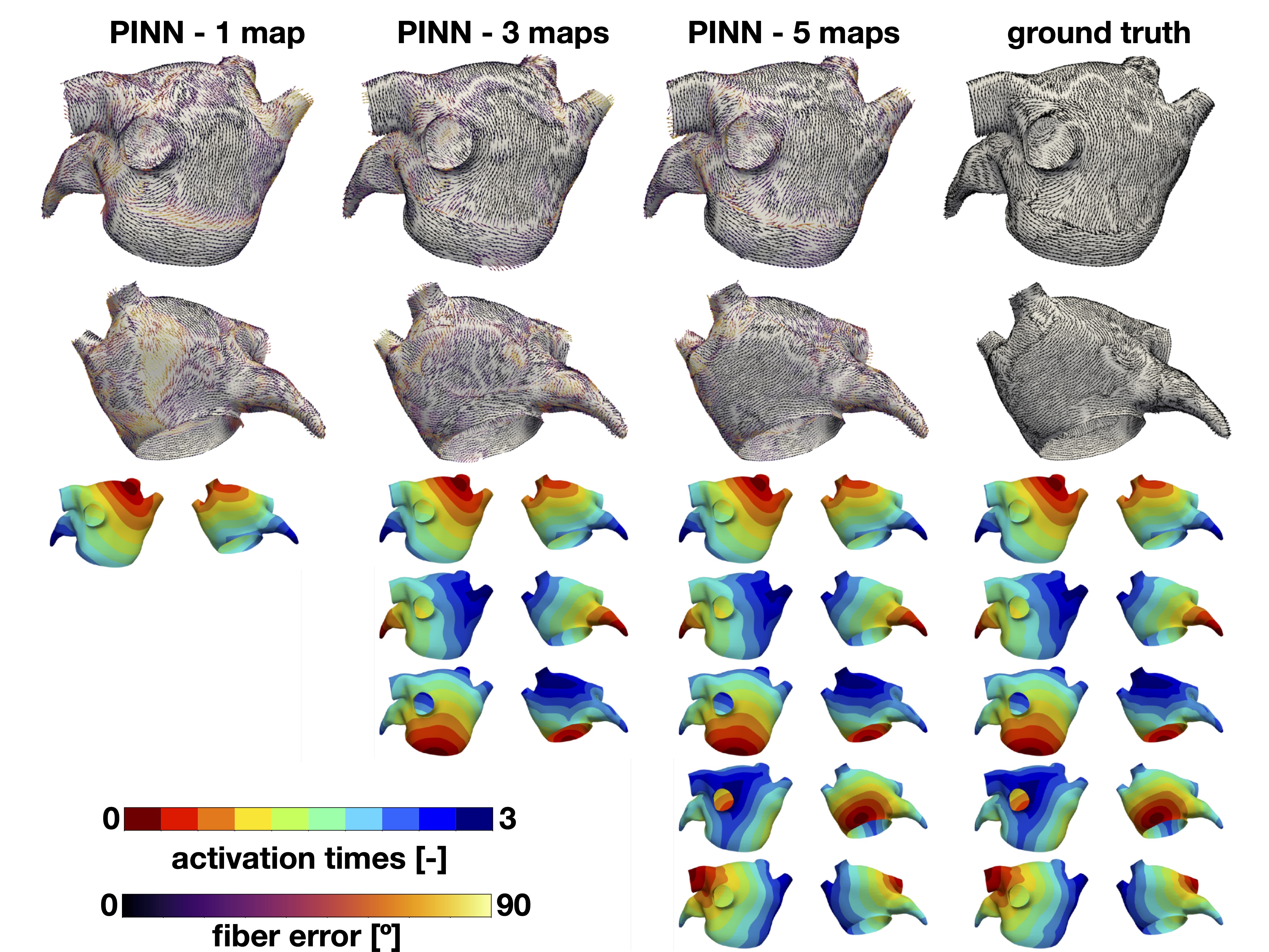}
    \caption{Predicted fiber orientations and activation times learned with 1, 3 and 5 maps. We show the results for the run with median error and hyper-parameters $\{\alpha_m:10^{-4},\ \alpha_e:10^{-5}\}$.}
    \label{fig:3Dexamples}
\end{figure}

In this numerical experiment we consider a patient-specific geometry of the left atrium. Fibers have been semi-automatically assigned from histological in a previous work~\cite{gharaviri2020epicardial}, shown in Figure~\ref{fig:3Dexamples}. The reference longitudinal $v_l$ and transverse velocity $v_t$ are respectively set equal to $0.6\,\mbox{m/s}$ and $0.4\,\mbox{m/s}$. The smooth basis necessary for the cosine of the fiber angle $a(\vec{x})$ is obtained from the atrial basis provided in Appendix~\ref{apx:smooth} (see also Figure~\ref{fig:smoothbasis}).
With the reference conductivity tensor, we generate 5 activation maps using an eikonal solver~\cite{grandits_fast_2021}, by pacing at 5 different locations well apart from each other. The first pacing site is randomly placed, whereas the subsequent ones are obtained through the farthest point sampling approach.

We train physics-informed neural networks that are fed either with 1, 3, or 5 maps. We use 870 measurement points that are split between the different cases, such that the total number of points remains constant.{The density ranges 9.9 samples/cm$^2$ when we use 1 map to 1.9 samples/cm$^2$. We set the hyperparameter to $\alpha_m:10^{-4}$}. We use neural networks of 7 hidden layers with 20 neurons to approximate $\hat{\phi}$ and one neural network with 5 hidden layers with 20 neurons each to generate $\hat{\ten{D}}$. Each network is then trained for 30\,000 Adam mini-batch iterations with batch-size 32. We also perform a sensitivity test for the regularization of conduction velocities and vary the parameter $\alpha_e$ between $10^{-9}$ and $10^{-3}$.

\begin{figure}[ht]
    \centering
    \includegraphics[width = \textwidth]{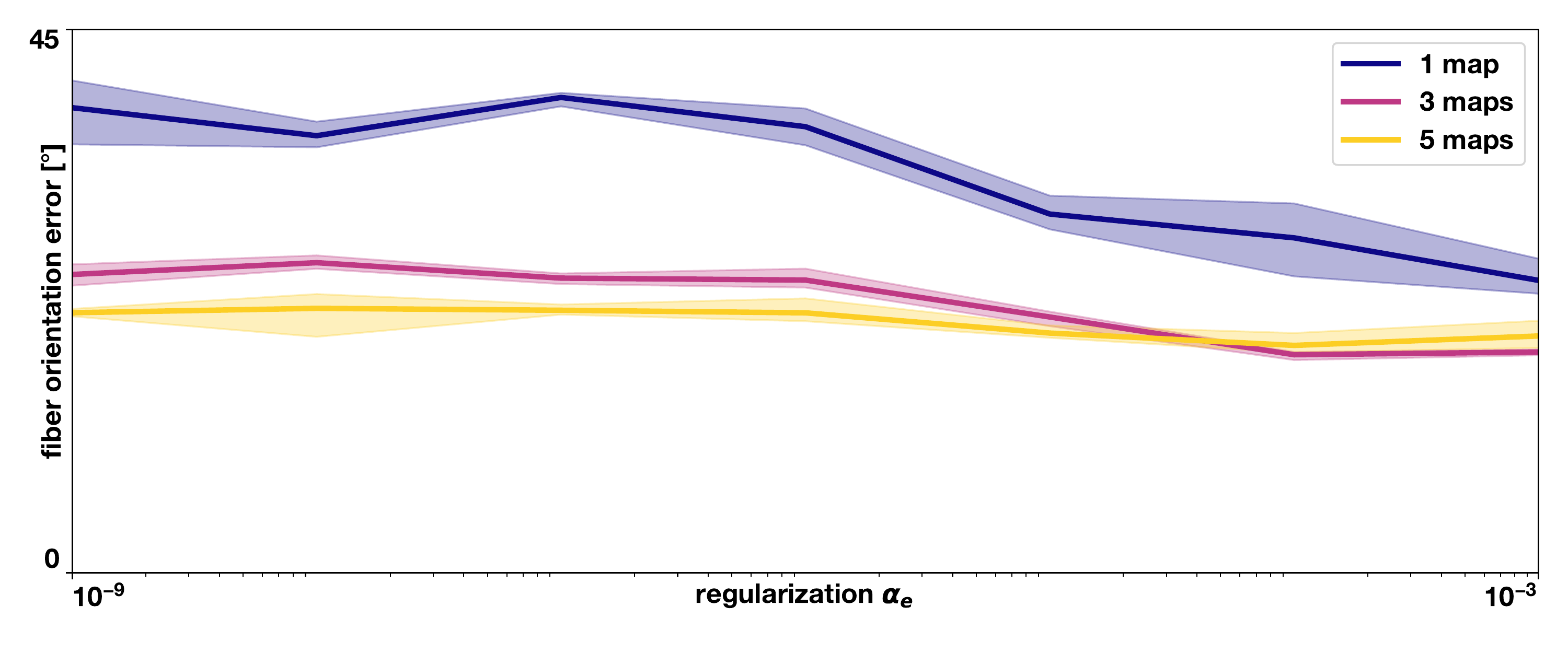}
    \caption{Fiber orientation error for $\alpha_m = 10^{-4}$ and $\alpha_{a} = 10^{-9}$. Error values of all runs (shaded area) and averages for 5 runs (solid lines) in 3D atrial geometry case with 1, 3 \& 5 maps as inputs using different values of the $\alpha_{e}$ conduction velocity total variation penalization weight.}
    \label{fig:3dtest}
\end{figure}

The results of the effect of conduction velocity regularization on the fiber error in this case are presented in Figure~\ref{fig:3dtest}. First, we observe that the fiber error is considerably higher when one map is used, compared to the cases with 3 and 5 maps. We see that increasing the regularization tends to decrease the error for all cases. Nonetheless, the case with one map is more sensitive to the amount of regularization applied than the cases of 3 and 5 maps. As in the 2D case, we see that increasing from 1 to 3 maps has a much bigger effect on fiber error than going from 3 to 5. We also note that the variability in the results decreases for all cases compared to the 2D case.

We show an example of these results with the case of median error and $\alpha_e: 10^{-5}$ in Figure~\ref{fig:3Dexamples}. In the right column, we can see the ground truth fiber distribution and the five maps fed to the different models. In general, we observe that all the maps are well reconstructed. However, we see regions of large fiber errors when training with one map. We see that the cases trained with 3 and 5 maps tend to concentrate the errors around the zones where there are sharp transitions in fiber directions, such as near the mitral annulus. Other regions of high error for all cases are near the pulmonary veins, which have a circumferential fiber orientation, which is hard to capture, most likely due to a lack of data. 

\section{Atrial geometries with diffusion tensor fibers}
\label{sec:dtmri}

\begin{figure}[ht]
    \centering
    \includegraphics[width = \textwidth]{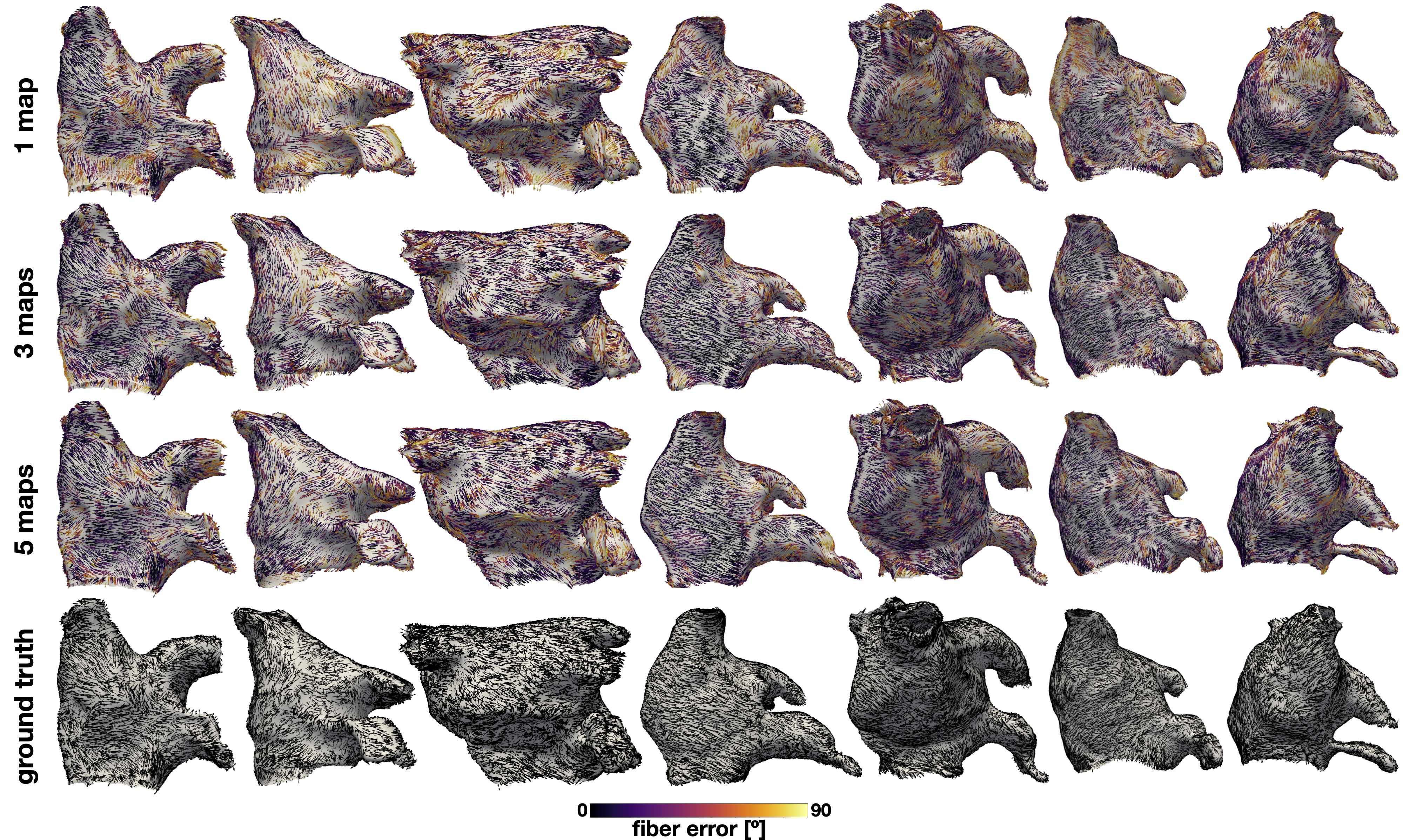}
    \caption{\textbf{Fiber predictions on diffusion tensor fibers.} We show the results for 7 different cases when we trained FiberNet with 1, 3 and 5 maps and no noise in the activation time data. {Errors in the cases with noise can be found in Table~\ref{tab:atlaserrors}. All the views are anteroposterior.}}
    \label{fig:atlas}
\end{figure}

For the next numerical experiment, we obtain fiber distributions for 7 different left atrium geometries from diffusion tensor magnetic resonance imaging \cite{roney2021constructing,pashakhanloo2016myofiber}. The images were taken \emph{ex-vivo} with an isotropic resolution of 0.4 mm and mapped to a biatrial mesh (Figure~\ref{fig:atlas}, bottom row). Here, we attempt to learn the endocardial fibers, generating synthetic activation maps by solving the eikonal equation as described in the previous sections with these fibers. We randomly select 1020 points on each of the surfaces and split them between 1, 3, and 5 maps to obtain the activation time data for training. {Since the different geometries have different surface areas, this numbers of measurements corresponds to densities that range from 9.06 to 18.68 samples/cm$^2$ for one map, and 1.81 to 3.73 samples/cm$^2$ when we used 5 maps.} In this experiment, we also test the robustness to noise of our method. We perturb the generated activation maps with Gaussian noise with zero mean and standard deviation of 0.1, 1 and 2 ms. We also include an additional validation step to test the overall performance of the method. We take the predicted and ground truth fibers and generate an additional map that is initiated from an activation site that is different from the maps used for training. Then, we compare these two maps and compute the error. In this way, we quantify how predictive are the learned fibers to model different scenarios, not included in the original dataset. For all cases, we set the hyper-parameters to  $\{\alpha_m:10^{-4},\ \alpha_e:10^{-5}\}$, which are the same used in the previous section.

\begin{figure}[ht]
    \centering
    \includegraphics[width = \textwidth]{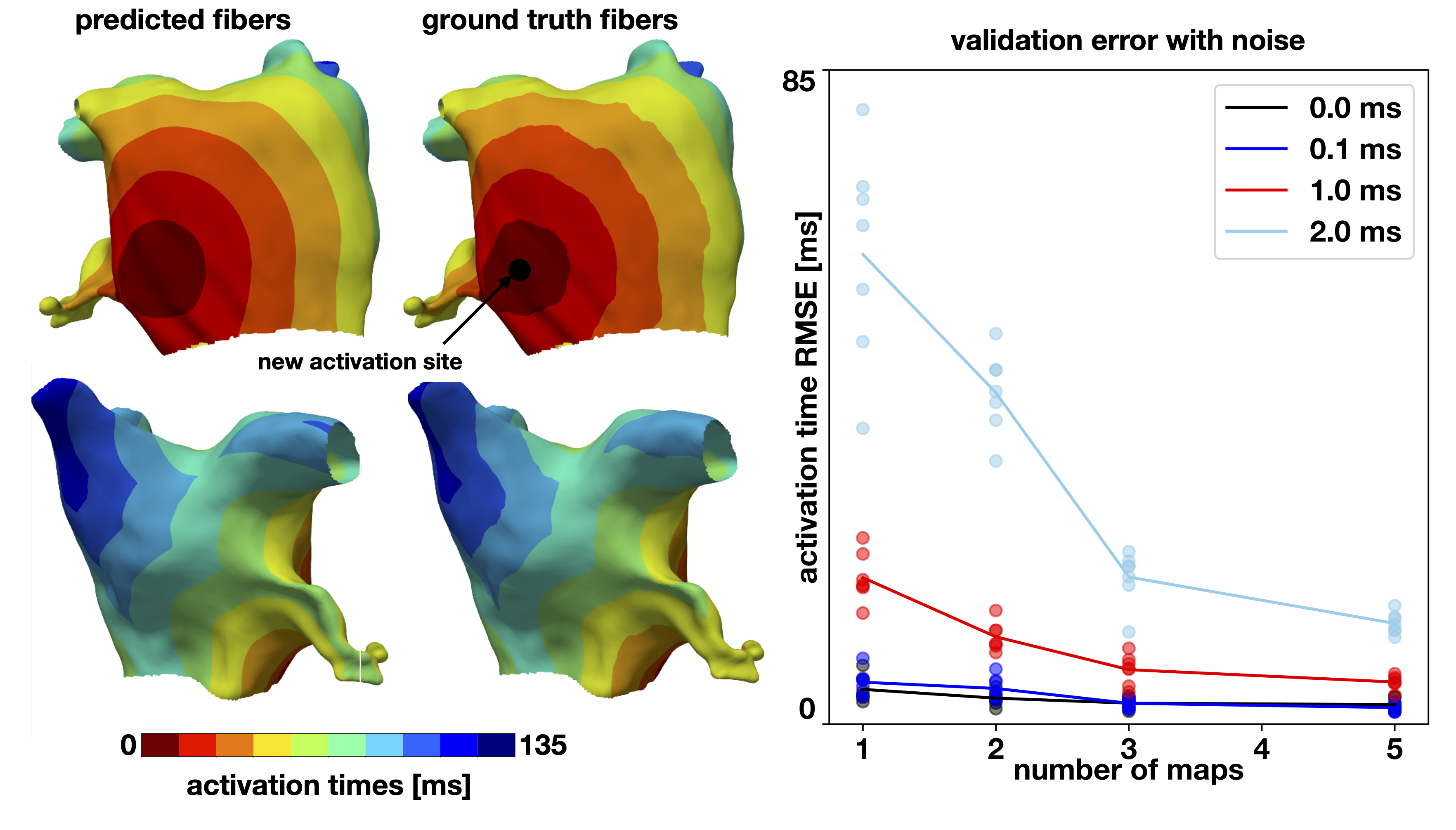}
    \caption{\textbf{Validation of conduction velocity tensor predictions with an additional map.} Left panel, an example of one the cases tested, where we create an activation from a point unseen during training. {The top row shows posteroanterior views and the bottom row shows anteroposterior views.} Right panel, we show a summary of the results of this validation for different number of maps and different levels of noise.}
    \label{fig:validation}
\end{figure}

The results of these experiments are presented in Figure~\ref{fig:atlas} and~\ref{fig:validation}, and Table~\ref{tab:atlaserrors}. In Figure~\ref{fig:atlas}, we show the fiber predictions and errors for the 7 different cases. We first note that the fibers obtained from {DT-MR imaging} are not smooth, with recurrent abrupt changes in direction. We also note that the approximation of the fibers improves as the number of maps is increased, which can be noted in the Figure~\ref{fig:atlas}, as the high error regions (in yellow) are less frequent for the cases with more maps. This qualitative result is confirmed in Table~\ref{tab:atlaserrors}, where the median fiber errors when training with one map range between 24.8$^\circ$ and 30.2$^\circ$, and they are decreased to the range of 18.3$^\circ$ and 23.2$^\circ$ when training with 3 maps, and the range of 16.2$^\circ$ and 23.3$^\circ$ when training with 5 maps. When we add noise to the activation measurements, we observe in the case of 0.1 ms that there is no clear trend, as some of the cases tend to improve and some cases tend to worsen their accuracy. Nonetheless, the variations in fiber errors are small, less than 5$^\circ$ for all cases. When we add 1 ms of noise, the performance of the method decreases for all cases and the number of maps. The fiber error increases, on average, 13.2$^\circ$ when training with one map, 12.4$^\circ$ when training with 3 maps, and 7.5$^\circ$ when training with 5 maps. {When we add 2 ms of noise, the error increases further, on average, to 61.0$^\circ$ with 1 map, 19.1$^\circ$ with 3 maps and 13.1$^\circ$ with 5 maps.} We note here that the hyper-parameters used for the noise study where calibrated for a noiseless case with synthetic geometry and fibers. Thus, it is expected that tuning these parameters, especially $\alpha_m$, which controls the relevance of the eikonal equation, might improve the results. Overall, we see from the trend that for a fixed amount of data points it is always better to distribute them in different maps, especially in the presence of noise. {The decrease in error is much more pronounced when increasing from 1 to 3 maps than when increasing from 3 to maps.}

For the validation with an additional map, we observe in Figure~\ref{fig:validation} that in general, using the learned fibers can lead to an accurate prediction of an unseen activation pattern. In the left panel, we see an example using 3 maps for the noiseless case. The map created with the predicted fibers is a smoother version of the one created with ground truth fibers, as the original data presented significant spatial variations. For the cases with 0 and 0.1 ms of noise, we see that number of maps used for training does not influence the predictive capabilities of the methods. For the noiseless case, the root mean squared errors in activation time, on average, range from 2.7 ms when training with 5 maps to 4.4 ms when training with one map. However, when we inject 1 ms of noise, these differences are more pronounced, with an average root mean squared errors ranging from 5.4 ms when training with 5 maps to 17.7 ms when training with one map. 

\section{Patient-specific geometry with multiple maps}
\label{sec:patspec}

In this final Section, we report the application of the proposed methodology to a patient-specific case.

\subsection{Data acquisition and pre-processing}

Data from a single patient who underwent a pulmonary veins isolation (PVI) procedure has been collected at the Institute of Cardiocentro Ticino with oral and written informed consent for the investigation. The study has been performed in compliance with the Declaration of Helsinki. Data consisted of an electrophysiological study of the left atrium (LA) right before and after PVI performed with an electroanatomical mapping system (RHYTHMIA HDx, Boston Scientific, USA) equipped with an ultra-density mapping catheter (INTELLIMAP ORION, Boston Scientific, USA). A total of 3 electroanatomical maps have been acquired before ablation. Each map contained several electrical recordings of the extracellular voltage of the endocardial atrial wall, as summarized in Table~\ref{tab:eam}.

\begin{table}[htb]
\centering
\begin{tabular}{ccccc}
Map Number & Rhythm & Pacing Location & Acquired EGMs & Accepted EGMs \\
\hline
         1 & Paced  & CS Distal    & 12\,405 & 3\,663 (29.5\%) \\
         2 & Paced  & CS Proximal  &  2\,713 &    442 (16.3\%) \\
         3 & SR     & ---          &  1\,563 &    159 (10.2\%) \\
\hline
Total & & & 16\,681 & 4264 (25.6\%) \\
\end{tabular}
\caption{Summary of the electroanatomical mapping data of the patient-specific case. SR: Sinus Rhythm, CS: Coronary Sinus, EGMs: Electrograms.}
\label{tab:eam}
\end{table}

Each recording consisted of the 3-D location of the recording electrode and 700\,ms of unfiltered electric signal (sampling resolution is 953\,Hz). Up to 64 recordings could be collected simultaneously, with only a portion of them in contact with the wall. Unipolar and bipolar electrograms have been automatically aligned in time to the R peak of the surface 12-lead ECG, simultaneously recorded. Pre-processing of the maps consisted in excluding electrograms with (i) low unipolar amplitude ($<0.05$\,mV); (ii) poor contact as indicated by the system; (iii) inconsistent surface P wave; {(iv) discrepancy of $>20$\,ms between unipolar and bipolar activation time. The activation time was computed from the unipolar signal} as the steepest negative deflection in the signal after the application of zero-phase forward and reverse 4-th order Butterworth with a cutoff frequency of 120\,Hz. As reported in Table~\ref{tab:eam}, the pre-processing excluded roughly 75\% of the points, most of them because of low amplitude. The analysis has been performed with MATLAB version R2021a.

The left atrial anatomy has been obtained from the mapping system. The original triangular mesh has been re-meshed with fTetWild~\cite{fTetWild} using the default parameters, with a final median edge length of 2.1\,mm (5\,112 points). Finally, the 3-D location of the electrodes was projected onto the atrial surface. It is worth mentioning that the mesh is only required for sampling collocation points and visualization, as by itself the proposed method is \emph{mesh free}. 

\subsection{Results}

\begin{figure}[htb]
    \centering
    \includegraphics[width = \textwidth]{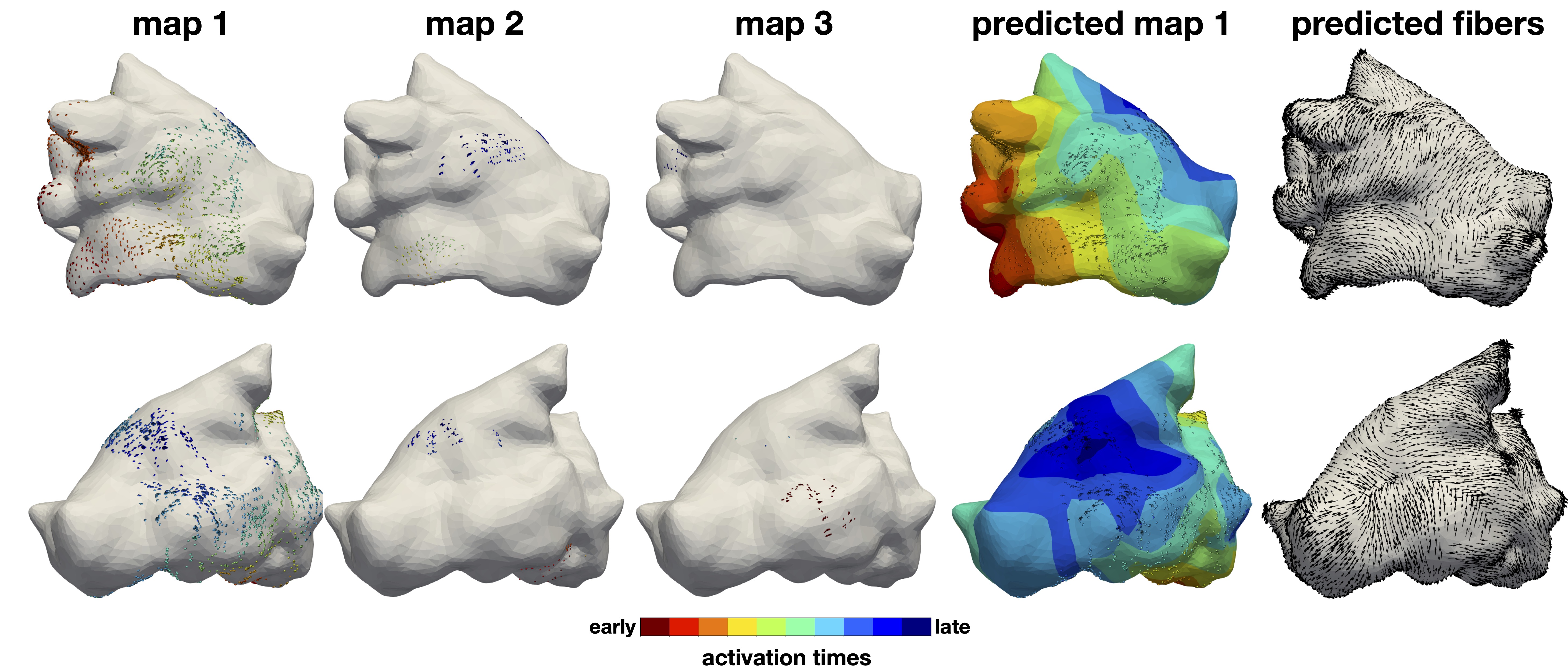}
    \caption{\textbf{Results of patient-specific case.} On the first 3 columns, we show the data points acquired colored by activation time, on the fourth column we show the learned map with the data points on top and on the last column we show the predicted fiber orientations.}
    \label{fig:patient}
\end{figure}

The pre-processed maps are shown in Figure~\ref{fig:patient}, the first 3 columns. We apply our method to these 3 maps and predict the fibers. We use the same hyper-parameters as in the previous sections. Although we do not have access to the real fiber orientations and we cannot compute the fiber error, we can check how well we have approximated the activation maps provided. Overall, we obtain the root mean squared error of 2.09 ms on the 3 activation maps. {We show the fit for map 1, which is the most complete, in Figure~\ref{fig:patient}, fourth column, demonstrating an excellent agreement with the measurements, which are shown as discrete points.} Finally, in the predicted fibers we observe some of the expected features of the left atrium fibers in Figure~\ref{fig:patient}, last column. We see that fibers go from the anterior to the posterior region through the atrial roof, and we also observe some regions where the fibers are aligned to the mitral annulus.

\section{Concluding remarks}
\label{sec:conclusion}

In this work, we present FiberNet, a novel method to estimate the complete conduction velocity tensor from a set of electroanatomical maps. The decomposition of the tensor simultaneously provides a patient-specific estimate of the fiber directions and conduction velocities. We tested our methodology in 2- and 3-dimensional synthetic examples, \textit{ex-vivo} fibers obtained from DT-MR imaging, and clinical data. Furthermore, we validated our approach by creating an additional map, unseen during training, and we are able to accurately predict the activation pattern. This step shows that the learned conductivity tensors can be used reliably for personalized modeling and digital twinning. From a methodological perspective, we showed that is not possible to estimate the conduction velocity tensors from a single map. In theory, at least 3 different maps are necessary, although having the correct regularization might reduce the error. {Each map should provide different propagation directions in the tissue, albeit not necessarily orthogonal.} In our numerical experiments, we see that for the same number of measurements, distributing them across different maps reduces the fiber error. Unsurprisingly, the error of estimating each of the maps increases, but this does not affect the accuracy of the fiber estimates.

FiberNet is currently limited to work on surfaces, which is a reasonable approximation for the atria under several circumstances~\cite{Roney2019UAC}.
We plan to extend FiberNet to work with solid bodies and apply it to the ventricles. {Transmurally, fibers rotate from an endocardial to an epicardial direction, and these directions may sensibly differ~\cite{streeter1969fiber}. In these cases, besides histological knowledge, we would require} additional data to learn the activation times transmurally, which might come from body surface potentials and prior histological knowledge~\cite{Bayer2012,streeter1969fiber,Piersanti2021Fibers}. Along this line, we are currently using minimally invasive data, which might limit the applicability of our approach. We plan to extend it to work with the surface electrocardiogram, by converting the activation times into extracellular potentials through the forward bidomain model~\cite{pezzuto2017evaluation,potse2018scalable,gillette2021framework}. Our method takes on the order of one hour to train. Even though this is an order of magnitude larger than for some of the local methods~\cite{Cantwell2015,roney2019technique}, this is a competitive time to solve an inverse problem~\cite{Grandits2021PIEMAP}. Furthermore, to our knowledge, this is the first global method to estimate conduction velocity tensor from multiple maps. We plan to accelerate the training process by transfer learning~\cite{lejeune2021exploring} and also by incorporating prior knowledge of the fiber distribution in the basis of vectors that we use to locally parametrize the surface. In this way, the functions that are approximated by the neural network are simpler and can be learned in fewer training steps due to spectral bias \cite{wang2021eigenvector}. {FiberNet cannot be applied during AF in its current formulation, because it assumes that all data points for a single map are from the same beat or, at least from the same activation pattern. However, it could be applied to a patch of tissue covered by the catheter, and it should perform comparably to a local method in estimating the CV. It is worth remarking that during AF, the CV varies beat to beat, due to the restitution properties of the tissue.}  Finally, FiberNet still needs further validation, {especially in its capability to extrapolate activation and fibers}, as we only tested it using either known fibers or known activation maps. {We plan to study and optimize the performance under noisy conditions to allow the application of this method in the clinical setting.}  {We plan to compare our approach to existing rule-based methods for generating fibers, as recently proposed~\cite{Wachter2015Fibers,Piersanti2021Fibers}. In the clinical case, it is unclear whether FiberNet estimated the endocardial fiber orientation, or rather a transmurally homogenized fiber orientation. The results likely depend on the degree of endo-epicardial dissociation~\cite{gharaviri2020epicardial}.} Ideally, we will acquire data from an animal model where we could obtain electroanatomical maps and \textit{ex-vivo} fiber orientations either from histological measurements or diffusion tensor imaging~\cite{Roney2021Fibers}.

FiberNet is a new step in the road of personalized medicine. We hope this method will aid the creation of patient-specific models for accurate diagnosis and procedure planning.

%


\backmatter

\bmhead{Acknowledgments}
This work was funded by an Open Seed Fund CORFO 14ENi2-26862 to CRH and FSC, the ANID -- Millennium Science Initiative Program -- NCN19-161 to FSC. We also acknowledge the School of Engineering computing cluster at Pontificia Universidad Católica de Chile for providing the computational resources for this study.
SP and FSC acknowledge the financial support of the Leading House for Latin American Region (Grant Agreement No.~RPG 2117). 
This work was also financially supported by the Theo Rossi di Montelera Foundation, the Metis Foundation Sergio Mantegazza, the Fidinam Foundation, the Horten Foundation to the Center for Computational Medicine in Cardiology. SP also acknowledges the CSCS-Swiss National Supercomputing Centre (Production Grant No.~s1074) and the Swiss Heart Foundation (No.~FF20042).
Finally, we are very grateful to Prof.~Angelo Auricchio and Dr.~Giulio Conte from Instituto Cardiocentro Ticino, EOC (Lugano, Switzerland) for providing the patient-specific data.

\bmhead{Compliance with Ethical Standards}

The authors have no relevant financial or non-financial interests to disclose. The clinical data used in this study was obtained with oral and written informed consent for the investigation. The study has been performed in compliance with the Declaration of Helsinki.

\bibliography{litra}


\begin{thebibliography}{53}
\ifx \bisbn   \undefined \def \bisbn  #1{ISBN #1}\fi
\ifx \binits  \undefined \def \binits#1{#1}\fi
\ifx \bauthor  \undefined \def \bauthor#1{#1}\fi
\ifx \batitle  \undefined \def \batitle#1{#1}\fi
\ifx \bjtitle  \undefined \def \bjtitle#1{#1}\fi
\ifx \bvolume  \undefined \def \bvolume#1{\textbf{#1}}\fi
\ifx \byear  \undefined \def \byear#1{#1}\fi
\ifx \bissue  \undefined \def \bissue#1{#1}\fi
\ifx \bfpage  \undefined \def \bfpage#1{#1}\fi
\ifx \blpage  \undefined \def \blpage #1{#1}\fi
\ifx \burl  \undefined \def \burl#1{\textsf{#1}}\fi
\ifx \doiurl  \undefined \def \doiurl#1{\url{https://doi.org/#1}}\fi
\ifx \betal  \undefined \def \betal{\textit{et al.}}\fi
\ifx \binstitute  \undefined \def \binstitute#1{#1}\fi
\ifx \binstitutionaled  \undefined \def \binstitutionaled#1{#1}\fi
\ifx \bctitle  \undefined \def \bctitle#1{#1}\fi
\ifx \beditor  \undefined \def \beditor#1{#1}\fi
\ifx \bpublisher  \undefined \def \bpublisher#1{#1}\fi
\ifx \bbtitle  \undefined \def \bbtitle#1{#1}\fi
\ifx \bedition  \undefined \def \bedition#1{#1}\fi
\ifx \bseriesno  \undefined \def \bseriesno#1{#1}\fi
\ifx \blocation  \undefined \def \blocation#1{#1}\fi
\ifx \bsertitle  \undefined \def \bsertitle#1{#1}\fi
\ifx \bsnm \undefined \def \bsnm#1{#1}\fi
\ifx \bsuffix \undefined \def \bsuffix#1{#1}\fi
\ifx \bparticle \undefined \def \bparticle#1{#1}\fi
\ifx \barticle \undefined \def \barticle#1{#1}\fi
\bibcommenthead
\ifx \bconfdate \undefined \def \bconfdate #1{#1}\fi
\ifx \botherref \undefined \def \botherref #1{#1}\fi
\ifx \url \undefined \def \url#1{\textsf{#1}}\fi
\ifx \bchapter \undefined \def \bchapter#1{#1}\fi
\ifx \bbook \undefined \def \bbook#1{#1}\fi
\ifx \bcomment \undefined \def \bcomment#1{#1}\fi
\ifx \oauthor \undefined \def \oauthor#1{#1}\fi
\ifx \citeauthoryear \undefined \def \citeauthoryear#1{#1}\fi
\ifx \endbibitem  \undefined \def \endbibitem {}\fi
\ifx \bconflocation  \undefined \def \bconflocation#1{#1}\fi
\ifx \arxivurl  \undefined \def \arxivurl#1{\textsf{#1}}\fi
\csname PreBibitemsHook\endcsname

\bibitem{corral_acero_digital_2020}
\begin{barticle}
\bauthor{\bsnm{Corral-Acero}, \binits{J.}},
\bauthor{\bsnm{Margara}, \binits{F.}},
\bauthor{\bsnm{Marciniak}, \binits{M.}},
\bauthor{\bsnm{Rodero}, \binits{C.}},
\bauthor{\bsnm{Loncaric}, \binits{F.}},
\bauthor{\bsnm{Feng}, \binits{Y.}},
\bauthor{\bsnm{Gilbert}, \binits{A.}},
\bauthor{\bsnm{Fernandes}, \binits{J.F.}},
\bauthor{\bsnm{Bukhari}, \binits{H.A.}},
\bauthor{\bsnm{Wajdan}, \binits{A.}},
\bauthor{\bsnm{Martinez}, \binits{M.V.}},
\bauthor{\bsnm{Santos}, \binits{M.S.}},
\bauthor{\bsnm{Shamohammdi}, \binits{M.}},
\bauthor{\bsnm{Luo}, \binits{H.}},
\bauthor{\bsnm{Westphal}, \binits{P.}},
\bauthor{\bsnm{Leeson}, \binits{P.}},
\bauthor{\bsnm{DiAchille}, \binits{P.}},
\bauthor{\bsnm{Gurev}, \binits{V.}},
\bauthor{\bsnm{Mayr}, \binits{M.}},
\bauthor{\bsnm{Geris}, \binits{L.}},
\bauthor{\bsnm{Pathmanathan}, \binits{P.}},
\bauthor{\bsnm{Morrison}, \binits{T.}},
\bauthor{\bsnm{Cornelussen}, \binits{R.}},
\bauthor{\bsnm{Prinzen}, \binits{F.}},
\bauthor{\bsnm{Delhaas}, \binits{T.}},
\bauthor{\bsnm{Doltra}, \binits{A.}},
\bauthor{\bsnm{Sitges}, \binits{M.}},
\bauthor{\bsnm{Vigmond}, \binits{E.J.}},
\bauthor{\bsnm{Zacur}, \binits{E.}},
\bauthor{\bsnm{Grau}, \binits{V.}},
\bauthor{\bsnm{Rodriguez}, \binits{B.}},
\bauthor{\bsnm{Remme}, \binits{E.W.}},
\bauthor{\bsnm{Niederer}, \binits{S.}},
\bauthor{\bsnm{Mortier}, \binits{P.}},
\bauthor{\bsnm{McLeod}, \binits{K.}},
\bauthor{\bsnm{Potse}, \binits{M.}},
\bauthor{\bsnm{Pueyo}, \binits{E.}},
\bauthor{\bsnm{Bueno-Orovio}, \binits{A.}},
\bauthor{\bsnm{Lamata}, \binits{P.}}:
\batitle{The ‘{Digital} {Twin}’ to enable the vision of precision
  cardiology}.
\bjtitle{European Heart Journal}
\bvolume{41}(\bissue{48}),
\bfpage{4556}--\blpage{4564}
(\byear{2020}).
\doiurl{10.1093/eurheartj/ehaa159}.
Accessed 2021-04-18
\end{barticle}
\endbibitem

\bibitem{peirlinck2021precision}
\begin{botherref}
\oauthor{\bsnm{Peirlinck}, \binits{M.}},
\oauthor{\bsnm{Costabal}, \binits{F.S.}},
\oauthor{\bsnm{Yao}, \binits{J.}},
\oauthor{\bsnm{Guccione}, \binits{J.}},
\oauthor{\bsnm{Tripathy}, \binits{S.}},
\oauthor{\bsnm{Wang}, \binits{Y.}},
\oauthor{\bsnm{Ozturk}, \binits{D.}},
\oauthor{\bsnm{Segars}, \binits{P.}},
\oauthor{\bsnm{Morrison}, \binits{T.}},
\oauthor{\bsnm{Levine}, \binits{S.}}, et al.:
Precision medicine in human heart modeling.
Biomechanics and modeling in mechanobiology,
1--29
(2021)
\end{botherref}
\endbibitem

\bibitem{o1986statistical}
\begin{botherref}
\oauthor{\bsnm{O'Sullivan}, \binits{F.}}:
A statistical perspective on ill-posed inverse problems.
Statistical science,
502--518
(1986)
\end{botherref}
\endbibitem

\bibitem{ho2009importance}
\begin{barticle}
\bauthor{\bsnm{Ho}, \binits{S.}},
\bauthor{\bsnm{Sanchez-Quintana}, \binits{D.}}:
\batitle{The importance of atrial structure and fibers}.
\bjtitle{Clinical Anatomy: The Official Journal of the American Association of
  Clinical Anatomists and the British Association of Clinical Anatomists}
\bvolume{22}(\bissue{1}),
\bfpage{52}--\blpage{63}
(\byear{2009})
\end{barticle}
\endbibitem

\bibitem{clerc1976directional}
\begin{barticle}
\bauthor{\bsnm{Clerc}, \binits{L.}}:
\batitle{Directional differences of impulse spread in trabecular muscle from
  mammalian heart.}
\bjtitle{The Journal of physiology}
\bvolume{255}(\bissue{2}),
\bfpage{335}--\blpage{346}
(\byear{1976})
\end{barticle}
\endbibitem

\bibitem{kotadia2020anisotropic}
\begin{barticle}
\bauthor{\bsnm{Kotadia}, \binits{I.}},
\bauthor{\bsnm{Whitaker}, \binits{J.}},
\bauthor{\bsnm{Roney}, \binits{C.}},
\bauthor{\bsnm{Niederer}, \binits{S.}},
\bauthor{\bsnm{O’Neill}, \binits{M.}},
\bauthor{\bsnm{Bishop}, \binits{M.}},
\bauthor{\bsnm{Wright}, \binits{M.}}:
\batitle{Anisotropic cardiac conduction}.
\bjtitle{Arrhythmia \& Electrophysiology Review}
\bvolume{9}(\bissue{4}),
\bfpage{202}
(\byear{2020})
\end{barticle}
\endbibitem

\bibitem{streeter1969fiber}
\begin{barticle}
\bauthor{\bsnm{Streeter~Jr}, \binits{D.D.}},
\bauthor{\bsnm{Spotnitz}, \binits{H.M.}},
\bauthor{\bsnm{Patel}, \binits{D.P.}},
\bauthor{\bsnm{Ross~Jr}, \binits{J.}},
\bauthor{\bsnm{Sonnenblick}, \binits{E.H.}}:
\batitle{Fiber orientation in the canine left ventricle during diastole and
  systole}.
\bjtitle{Circulation research}
\bvolume{24}(\bissue{3}),
\bfpage{339}--\blpage{347}
(\byear{1969})
\end{barticle}
\endbibitem

\bibitem{Bayer2012}
\begin{barticle}
\bauthor{\bsnm{Bayer}, \binits{J.D.}},
\bauthor{\bsnm{Blake}, \binits{R.C.}},
\bauthor{\bsnm{Plank}, \binits{G.}},
\bauthor{\bsnm{Trayanova}, \binits{N.a.}}:
\batitle{{A novel rule-based algorithm for assigning myocardial fiber
  orientation to computational heart models.}}
\bjtitle{Annals of biomedical engineering}
\bvolume{40}(\bissue{10}),
\bfpage{2243}--\blpage{54}
(\byear{2012}).
\doiurl{10.1007/s10439-012-0593-5}
\end{barticle}
\endbibitem

\bibitem{gonzales2013three}
\begin{barticle}
\bauthor{\bsnm{Gonzales}, \binits{M.J.}},
\bauthor{\bsnm{Sturgeon}, \binits{G.}},
\bauthor{\bsnm{Krishnamurthy}, \binits{A.}},
\bauthor{\bsnm{Hake}, \binits{J.}},
\bauthor{\bsnm{Jonas}, \binits{R.}},
\bauthor{\bsnm{Stark}, \binits{P.}},
\bauthor{\bsnm{Rappel}, \binits{W.-J.}},
\bauthor{\bsnm{Narayan}, \binits{S.M.}},
\bauthor{\bsnm{Zhang}, \binits{Y.}},
\bauthor{\bsnm{Segars}, \binits{W.P.}}, \betal:
\batitle{A three-dimensional finite element model of human atrial anatomy: new
  methods for cubic hermite meshes with extraordinary vertices}.
\bjtitle{Medical image analysis}
\bvolume{17}(\bissue{5}),
\bfpage{525}--\blpage{537}
(\byear{2013})
\end{barticle}
\endbibitem

\bibitem{Wachter2015Fibers}
\begin{barticle}
\bauthor{\bsnm{Wachter}, \binits{A.}},
\bauthor{\bsnm{Loewe}, \binits{A.}},
\bauthor{\bsnm{Krueger}, \binits{M.W.}},
\bauthor{\bsnm{Dössel}, \binits{O.}},
\bauthor{\bsnm{Seemann}, \binits{G.}}:
\batitle{Mesh structure-independent modeling of patient-specific atrial fiber
  orientation}.
\bjtitle{Current Directions in Biomedical Engineering}
\bvolume{1}(\bissue{1}),
\bfpage{409}--\blpage{412}
(\byear{2015}).
\doiurl{10.1515/cdbme-2015-0099}
\end{barticle}
\endbibitem

\bibitem{Roney2019UAC}
\begin{barticle}
\bauthor{\bsnm{Roney}, \binits{C.H.}},
\bauthor{\bsnm{Pashaei}, \binits{A.}},
\bauthor{\bsnm{Meo}, \binits{M.}},
\bauthor{\bsnm{Dubois}, \binits{R.}},
\bauthor{\bsnm{Boyle}, \binits{P.M.}},
\bauthor{\bsnm{Trayanova}, \binits{N.A.}},
\bauthor{\bsnm{Cochet}, \binits{H.}},
\bauthor{\bsnm{Niederer}, \binits{S.A.}},
\bauthor{\bsnm{Vigmond}, \binits{E.J.}}:
\batitle{Universal atrial coordinates applied to visualisation, registration
  and construction of patient specific meshes}.
\bjtitle{Medical Image Analysis}
\bvolume{55},
\bfpage{65}--\blpage{75}
(\byear{2019})
\end{barticle}
\endbibitem

\bibitem{roney2021constructing}
\begin{barticle}
\bauthor{\bsnm{Roney}, \binits{C.H.}},
\bauthor{\bsnm{Bendikas}, \binits{R.}},
\bauthor{\bsnm{Pashakhanloo}, \binits{F.}},
\bauthor{\bsnm{Corrado}, \binits{C.}},
\bauthor{\bsnm{Vigmond}, \binits{E.J.}},
\bauthor{\bsnm{McVeigh}, \binits{E.R.}},
\bauthor{\bsnm{Trayanova}, \binits{N.A.}},
\bauthor{\bsnm{Niederer}, \binits{S.A.}}:
\batitle{Constructing a human atrial fibre atlas}.
\bjtitle{Annals of biomedical engineering}
\bvolume{49},
\bfpage{233}--\blpage{250}
(\byear{2021})
\end{barticle}
\endbibitem

\bibitem{Piersanti2021Fibers}
\begin{barticle}
\bauthor{\bsnm{Piersanti}, \binits{R.}},
\bauthor{\bsnm{Africa}, \binits{P.C.}},
\bauthor{\bsnm{Fedele}, \binits{M.}},
\bauthor{\bsnm{Vergara}, \binits{C.}},
\bauthor{\bsnm{Ded{\`e}}, \binits{L.}},
\bauthor{\bsnm{Corno}, \binits{A.F.}},
\bauthor{\bsnm{Quarteroni}, \binits{A.}}:
\batitle{Modeling cardiac muscle fibers in ventricular and atrial
  electrophysiology simulations}.
\bjtitle{Computer Methods in Applied Mechanics and Engineering}
\bvolume{373},
\bfpage{113468}
(\byear{2021})
\end{barticle}
\endbibitem

\bibitem{lombaert2012human}
\begin{barticle}
\bauthor{\bsnm{Lombaert}, \binits{H.}},
\bauthor{\bsnm{Peyrat}, \binits{J.-M.}},
\bauthor{\bsnm{Croisille}, \binits{P.}},
\bauthor{\bsnm{Rapacchi}, \binits{S.}},
\bauthor{\bsnm{Fanton}, \binits{L.}},
\bauthor{\bsnm{Cheriet}, \binits{F.}},
\bauthor{\bsnm{Clarysse}, \binits{P.}},
\bauthor{\bsnm{Magnin}, \binits{I.}},
\bauthor{\bsnm{Delingette}, \binits{H.}},
\bauthor{\bsnm{Ayache}, \binits{N.}}:
\batitle{Human atlas of the cardiac fiber architecture: study on a healthy
  population}.
\bjtitle{IEEE transactions on medical imaging}
\bvolume{31}(\bissue{7}),
\bfpage{1436}--\blpage{1447}
(\byear{2012})
\end{barticle}
\endbibitem

\bibitem{teh2016resolving}
\begin{barticle}
\bauthor{\bsnm{Teh}, \binits{I.}},
\bauthor{\bsnm{McClymont}, \binits{D.}},
\bauthor{\bsnm{Burton}, \binits{R.A.}},
\bauthor{\bsnm{Maguire}, \binits{M.L.}},
\bauthor{\bsnm{Whittington}, \binits{H.J.}},
\bauthor{\bsnm{Lygate}, \binits{C.A.}},
\bauthor{\bsnm{Kohl}, \binits{P.}},
\bauthor{\bsnm{Schneider}, \binits{J.E.}}:
\batitle{Resolving fine cardiac structures in rats with high-resolution
  diffusion tensor imaging}.
\bjtitle{Scientific reports}
\bvolume{6}(\bissue{1}),
\bfpage{1}--\blpage{14}
(\byear{2016})
\end{barticle}
\endbibitem

\bibitem{pashakhanloo2016myofiber}
\begin{barticle}
\bauthor{\bsnm{Pashakhanloo}, \binits{F.}},
\bauthor{\bsnm{Herzka}, \binits{D.A.}},
\bauthor{\bsnm{Ashikaga}, \binits{H.}},
\bauthor{\bsnm{Mori}, \binits{S.}},
\bauthor{\bsnm{Gai}, \binits{N.}},
\bauthor{\bsnm{Bluemke}, \binits{D.A.}},
\bauthor{\bsnm{Trayanova}, \binits{N.A.}},
\bauthor{\bsnm{McVeigh}, \binits{E.R.}}:
\batitle{Myofiber architecture of the human atria as revealed by submillimeter
  diffusion tensor imaging}.
\bjtitle{Circulation: arrhythmia and electrophysiology}
\bvolume{9}(\bissue{4}),
\bfpage{004133}
(\byear{2016})
\end{barticle}
\endbibitem

\bibitem{Cantwell2015}
\begin{barticle}
\bauthor{\bsnm{Cantwell}, \binits{C.D.}},
\bauthor{\bsnm{Roney}, \binits{C.H.}},
\bauthor{\bsnm{Ng}, \binits{F.S.}},
\bauthor{\bsnm{Siggers}, \binits{J.H.}},
\bauthor{\bsnm{Sherwin}, \binits{S.J.}},
\bauthor{\bsnm{Peters}, \binits{N.S.}}:
\batitle{{Techniques for automated local activation time annotation and
  conduction velocity estimation in cardiac mapping}}.
\bjtitle{Computers in Biology and Medicine}
\bvolume{65},
\bfpage{229}--\blpage{242}
(\byear{2015}).
\doiurl{10.1016/j.compbiomed.2015.04.027}
\end{barticle}
\endbibitem

\bibitem{Coveney2020GP}
\begin{barticle}
\bauthor{\bsnm{Coveney}, \binits{S.}},
\bauthor{\bsnm{Corrado}, \binits{C.}},
\bauthor{\bsnm{Roney}, \binits{C.H.}},
\bauthor{\bsnm{O'Hare}, \binits{D.}},
\bauthor{\bsnm{Williams}, \binits{S.E.}},
\bauthor{\bsnm{O'Neill}, \binits{M.D.}},
\bauthor{\bsnm{Niederer}, \binits{S.A.}},
\bauthor{\bsnm{Clayton}, \binits{R.H.}},
\bauthor{\bsnm{Oakley}, \binits{J.E.}},
\bauthor{\bsnm{Wilkinson}, \binits{R.D.}}:
\batitle{Gaussian process manifold interpolation for probabilistic atrial
  activation maps and uncertain conduction velocity}.
\bjtitle{Philosophical Transactions of the Royal Society A: Mathematical,
  Physical and Engineering Sciences}
\bvolume{378}(\bissue{2173}),
\bfpage{20190345}
(\byear{2020})
\end{barticle}
\endbibitem

\bibitem{roney2019technique}
\begin{barticle}
\bauthor{\bsnm{Roney}, \binits{C.H.}},
\bauthor{\bsnm{Whitaker}, \binits{J.}},
\bauthor{\bsnm{Sim}, \binits{I.}},
\bauthor{\bsnm{O'Neill}, \binits{L.}},
\bauthor{\bsnm{Mukherjee}, \binits{R.K.}},
\bauthor{\bsnm{Razeghi}, \binits{O.}},
\bauthor{\bsnm{Vigmond}, \binits{E.J.}},
\bauthor{\bsnm{Wright}, \binits{M.}},
\bauthor{\bsnm{O'Neill}, \binits{M.D.}},
\bauthor{\bsnm{Williams}, \binits{S.E.}}, \betal:
\batitle{A technique for measuring anisotropy in atrial conduction to estimate
  conduction velocity and atrial fibre direction}.
\bjtitle{Computers in biology and medicine}
\bvolume{104},
\bfpage{278}--\blpage{290}
(\byear{2019})
\end{barticle}
\endbibitem

\bibitem{franzone1990wavefront}
\begin{barticle}
\bauthor{\bsnm{Colli~Franzone}, \binits{P.}},
\bauthor{\bsnm{Guerri}, \binits{L.}},
\bauthor{\bsnm{Rovida}, \binits{S.}}:
\batitle{Wavefront propagation in an activation model of the anisotropic
  cardiac tissue: asymptotic analysis and numerical simulations}.
\bjtitle{Journal of mathematical biology}
\bvolume{28}(\bissue{2}),
\bfpage{121}--\blpage{176}
(\byear{1990})
\end{barticle}
\endbibitem

\bibitem{Grandits2021PIEMAP}
\begin{bchapter}
\bauthor{\bsnm{Grandits}, \binits{T.}},
\bauthor{\bsnm{Pezzuto}, \binits{S.}},
\bauthor{\bsnm{Lubrecht}, \binits{J.M.}},
\bauthor{\bsnm{Pock}, \binits{T.}},
\bauthor{\bsnm{Plank}, \binits{G.}},
\bauthor{\bsnm{Krause}, \binits{R.}}:
\bctitle{{PIEMAP}: Personalized inverse eikonal model from cardiac
  electro-anatomical maps}.
In: \beditor{\bsnm{Puyol~Anton}, \binits{E.}},
\beditor{\bsnm{Pop}, \binits{M.}},
\beditor{\bsnm{Sermesant}, \binits{M.}},
\beditor{\bsnm{Campello}, \binits{V.}},
\beditor{\bsnm{Lalande}, \binits{A.}},
\beditor{\bsnm{Lekadir}, \binits{K.}},
\beditor{\bsnm{Suinesiaputra}, \binits{A.}},
\beditor{\bsnm{Camara}, \binits{O.}},
\beditor{\bsnm{Young}, \binits{A.}} (eds.)
\bbtitle{STACOM. M{\&}Ms and EMIDEC Challenges}.
\bsertitle{Lecture Notes in Computer Science},
vol. \bseriesno{12592},
pp. \bfpage{76}--\blpage{86}.
\bpublisher{Springer},
\blocation{Cham}
(\byear{2021})
\end{bchapter}
\endbibitem

\bibitem{Lubrecht2021PIEMAP}
\begin{barticle}
\bauthor{\bsnm{Lubrecht}, \binits{J.M.}},
\bauthor{\bsnm{Grandits}, \binits{T.}},
\bauthor{\bsnm{Gharaviri}, \binits{A.}},
\bauthor{\bsnm{Schotten}, \binits{U.}},
\bauthor{\bsnm{Pock}, \binits{T.}},
\bauthor{\bsnm{Plank}, \binits{G.}},
\bauthor{\bsnm{Krause}, \binits{R.}},
\bauthor{\bsnm{Auricchio}, \binits{A.}},
\bauthor{\bsnm{Pezzuto}, \binits{S.}}:
\batitle{Automatic reconstruction of the left atrium activation from sparse
  intracardiac contact recordings by inverse estimate of fiber structure and
  anisotropic conduction in a patient-specific model}.
\bjtitle{EP Europace}
\bvolume{23},
\bfpage{63}--\blpage{70}
(\byear{2021})
\end{barticle}
\endbibitem

\bibitem{raissi_pinn_2019}
\begin{barticle}
\bauthor{\bsnm{Raissi}, \binits{M.}},
\bauthor{\bsnm{Perdikaris}, \binits{P.}},
\bauthor{\bsnm{Karniadakis}, \binits{G.E.}}:
\batitle{Physics-informed neural networks: {A} deep learning framework for
  solving forward and inverse problems involving nonlinear partial differential
  equations}.
\bjtitle{J.~Comp.~Phys.}
\bvolume{378},
\bfpage{686}--\blpage{707}
(\byear{2019})
\end{barticle}
\endbibitem

\bibitem{grandits_learning_2021}
\begin{bchapter}
\bauthor{\bsnm{Grandits}, \binits{T.}},
\bauthor{\bsnm{Pezzuto}, \binits{S.}},
\bauthor{\bsnm{Sahli~Costabal}, \binits{F.}},
\bauthor{\bsnm{Perdikaris}, \binits{P.}},
\bauthor{\bsnm{Pock}, \binits{T.}},
\bauthor{\bsnm{Plank}, \binits{G.}},
\bauthor{\bsnm{Krause}, \binits{R.}}:
\bctitle{Learning atrial fiber orientations and conductivity tensors from
  intracardiac maps using physics-informed neural network}.
In: \beditor{\bsnm{Ennis}, \binits{D.B.}},
\beditor{\bsnm{Perotti}, \binits{L.E.}},
\beditor{\bsnm{Wang}, \binits{V.Y.}} (eds.)
\bbtitle{Functional Imaging and Modeling of the Heart. FIMH 2021}.
\bsertitle{Lecture Notes in Computer Science},
vol. \bseriesno{12738},
pp. \bfpage{650}--\blpage{658}.
\bpublisher{Springer},
\blocation{Cham}
(\byear{2021}).
\doiurl{978-3-030-78710-3\_62}
\end{bchapter}
\endbibitem

\bibitem{roneyAtlas}
\begin{botherref}
\oauthor{\bsnm{Roney}, \binits{C.}},
\oauthor{\bsnm{Bendikas}, \binits{R.}},
\oauthor{\bsnm{Pashakhanloo}, \binits{F.}},
\oauthor{\bsnm{Corrado}, \binits{C.}},
\oauthor{\bsnm{Vigmond}, \binits{E.}},
\oauthor{\bsnm{McVeigh}, \binits{E.}},
\oauthor{\bsnm{Trayanova}, \binits{N.}},
\oauthor{\bsnm{Niederer}, \binits{S.}}:
{Constructing a Human Atrial Fibre Atlas}.
Zenodo
(2020).
\doiurl{10.5281/zenodo.3764917}
\end{botherref}
\endbibitem

\bibitem{fu_fast_2011}
\begin{barticle}
\bauthor{\bsnm{Fu}, \binits{Z.}},
\bauthor{\bsnm{Jeong}, \binits{W.}},
\bauthor{\bsnm{Pan}, \binits{Y.}},
\bauthor{\bsnm{Kirby}, \binits{R.}},
\bauthor{\bsnm{Whitaker}, \binits{R.}}:
\batitle{A {Fast} {Iterative} {Method} for {Solving} the {Eikonal} {Equation}
  on {Triangulated} {Surfaces}}.
\bjtitle{SIAM Journal on Scientific Computing}
\bvolume{33}(\bissue{5}),
\bfpage{2468}--\blpage{2488}
(\byear{2011}).
\doiurl{10.1137/100788951}.
Accessed 2018-07-26
\end{barticle}
\endbibitem

\bibitem{grandits_fast_2021}
\begin{barticle}
\bauthor{\bsnm{Grandits}, \binits{T.}}:
\batitle{A {Fast} {Iterative} {Method} {Python} package}.
\bjtitle{Journal of Open Source Software}
\bvolume{6}(\bissue{66}),
\bfpage{3641}
(\byear{2021}).
\doiurl{10.21105/joss.03641}.
Accessed 2021-10-23
\end{barticle}
\endbibitem

\bibitem{Verma2018CV}
\begin{barticle}
\bauthor{\bsnm{Verma}, \binits{B.}},
\bauthor{\bsnm{Oesterlein}, \binits{T.}},
\bauthor{\bsnm{Loewe}, \binits{A.}},
\bauthor{\bsnm{Luik}, \binits{A.}},
\bauthor{\bsnm{Schmitt}, \binits{C.}},
\bauthor{\bsnm{D{\"o}ssel}, \binits{O.}}:
\batitle{Regional conduction velocity calculation from clinical multichannel
  electrograms in human atria}.
\bjtitle{Computers in Biology and Medicine}
\bvolume{92},
\bfpage{188}--\blpage{196}
(\byear{2018}).
\doiurl{10.1016/j.compbiomed.2017.11.017}
\end{barticle}
\endbibitem

\bibitem{vanSchie2021CV}
\begin{barticle}
\bauthor{\bparticle{van} \bsnm{Schie}, \binits{M.S.}},
\bauthor{\bsnm{Heida}, \binits{A.}},
\bauthor{\bsnm{Taverne}, \binits{Y.J.H.J.}},
\bauthor{\bsnm{Bogers}, \binits{A.J.J.C.}},
\bauthor{\bparticle{de} \bsnm{Groot}, \binits{N.M.S.}}:
\batitle{{Identification of local atrial conduction heterogeneities using
  high-density conduction velocity estimation}}.
\bjtitle{EP Europace}
\bvolume{23}(\bissue{11}),
\bfpage{1815}--\blpage{1825}
(\byear{2021})
\end{barticle}
\endbibitem

\bibitem{Pagani2021AF}
\begin{barticle}
\bauthor{\bsnm{Pagani}, \binits{S.}},
\bauthor{\bsnm{Dede'}, \binits{L.}},
\bauthor{\bsnm{Frontera}, \binits{A.}},
\bauthor{\bsnm{Salvador}, \binits{M.}},
\bauthor{\bsnm{Limite}, \binits{L.R.}},
\bauthor{\bsnm{Manzoni}, \binits{A.}},
\bauthor{\bsnm{Lipartiti}, \binits{F.}},
\bauthor{\bsnm{Tsitsinakis}, \binits{G.}},
\bauthor{\bsnm{Hadjis}, \binits{A.}},
\bauthor{\bsnm{Della~Bella}, \binits{P.}},
\bauthor{\bsnm{Quarteroni}, \binits{A.}}:
\batitle{A computational study of the electrophysiological substrate in
  patients suffering from atrial fibrillation}.
\bjtitle{Frontiers in Physiology}
\bvolume{12},
\bfpage{927}
(\byear{2021})
\end{barticle}
\endbibitem

\bibitem{Good2021CV}
\begin{barticle}
\bauthor{\bsnm{Good}, \binits{W.W.}},
\bauthor{\bsnm{Gillette}, \binits{K.K.}},
\bauthor{\bsnm{Zenger}, \binits{B.}},
\bauthor{\bsnm{Bergquist}, \binits{J.A.}},
\bauthor{\bsnm{Rupp}, \binits{L.C.}},
\bauthor{\bsnm{Tate}, \binits{J.}},
\bauthor{\bsnm{Anderson}, \binits{D.}},
\bauthor{\bsnm{Gsell}, \binits{M.A.F.}},
\bauthor{\bsnm{Plank}, \binits{G.}},
\bauthor{\bsnm{MacLeod}, \binits{R.S.}}:
\batitle{Estimation and validation of cardiac conduction velocity and wavefront
  reconstruction using epicardial and volumetric data}.
\bjtitle{IEEE Transactions on Biomedical Engineering}
\bvolume{68}(\bissue{11}),
\bfpage{3290}--\blpage{3300}
(\byear{2021}).
\doiurl{10.1109/TBME.2021.3069792}
\end{barticle}
\endbibitem

\bibitem{Nothstein2021CVAR}
\begin{barticle}
\bauthor{\bsnm{Nothstein}, \binits{M.}},
\bauthor{\bsnm{Luik}, \binits{A.}},
\bauthor{\bsnm{Jadidi}, \binits{A.}},
\bauthor{\bsnm{S{\'a}nchez}, \binits{J.}},
\bauthor{\bsnm{Unger}, \binits{L.A.}},
\bauthor{\bsnm{W{\"u}lfers}, \binits{E.M.}},
\bauthor{\bsnm{D{\"o}ssel}, \binits{O.}},
\bauthor{\bsnm{Seemann}, \binits{G.}},
\bauthor{\bsnm{Schmitt}, \binits{C.}},
\bauthor{\bsnm{Loewe}, \binits{A.}}:
\batitle{Cvar-seg: An automated signal segmentation pipeline for conduction
  velocity and amplitude restitution}.
\bjtitle{Frontiers in Physiology}
\bvolume{12},
\bfpage{746}
(\byear{2021}).
\doiurl{10.3389/fphys.2021.673047}
\end{barticle}
\endbibitem

\bibitem{sahli2020physics}
\begin{barticle}
\bauthor{\bsnm{Sahli~Costabal}, \binits{F.}},
\bauthor{\bsnm{Yang}, \binits{Y.}},
\bauthor{\bsnm{Perdikaris}, \binits{P.}},
\bauthor{\bsnm{Hurtado}, \binits{D.E.}},
\bauthor{\bsnm{Kuhl}, \binits{E.}}:
\batitle{Physics-informed neural networks for cardiac activation mapping}.
\bjtitle{Frontiers in Physics}
\bvolume{8},
\bfpage{42}
(\byear{2020})
\end{barticle}
\endbibitem

\bibitem{gander1994least}
\begin{barticle}
\bauthor{\bsnm{Gander}, \binits{W.}},
\bauthor{\bsnm{Golub}, \binits{G.H.}},
\bauthor{\bsnm{Strebel}, \binits{R.}}:
\batitle{Least-squares fitting of circles and ellipses}.
\bjtitle{BIT Numerical Mathematics}
\bvolume{34}(\bissue{4}),
\bfpage{558}--\blpage{578}
(\byear{1994})
\end{barticle}
\endbibitem

\bibitem{masse2016resolving}
\begin{barticle}
\bauthor{\bsnm{Mass{\'e}}, \binits{S.}},
\bauthor{\bsnm{Magtibay}, \binits{K.}},
\bauthor{\bsnm{Jackson}, \binits{N.}},
\bauthor{\bsnm{Asta}, \binits{J.}},
\bauthor{\bsnm{Kusha}, \binits{M.}},
\bauthor{\bsnm{Zhang}, \binits{B.}},
\bauthor{\bsnm{Balachandran}, \binits{R.}},
\bauthor{\bsnm{Radisic}, \binits{M.}},
\bauthor{\bsnm{Deno}, \binits{D.C.}},
\bauthor{\bsnm{Nanthakumar}, \binits{K.}}:
\batitle{Resolving myocardial activation with novel omnipolar electrograms}.
\bjtitle{Circulation: Arrhythmia and Electrophysiology}
\bvolume{9}(\bissue{7}),
\bfpage{004107}
(\byear{2016})
\end{barticle}
\endbibitem

\bibitem{Gaeta2021DELTA}
\begin{barticle}
\bauthor{\bsnm{Gaeta}, \binits{S.}},
\bauthor{\bsnm{Bahnson}, \binits{T.D.}},
\bauthor{\bsnm{Henriquez}, \binits{C.}}:
\batitle{High-resolution measurement of local activation time differences from
  bipolar electrogram amplitude}.
\bjtitle{Frontiers in Physiology}
\bvolume{12},
\bfpage{536}
(\byear{2021})
\end{barticle}
\endbibitem

\bibitem{yang2015estimation}
\begin{barticle}
\bauthor{\bsnm{Yang}, \binits{H.}},
\bauthor{\bsnm{Veneziani}, \binits{A.}}:
\batitle{Estimation of cardiac conductivities in ventricular tissue by a
  variational approach}.
\bjtitle{Inverse Problems}
\bvolume{31}(\bissue{11}),
\bfpage{115001}
(\byear{2015})
\end{barticle}
\endbibitem

\bibitem{Barone2020Conductivity}
\begin{barticle}
\bauthor{\bsnm{Barone}, \binits{A.}},
\bauthor{\bsnm{Gizzi}, \binits{A.}},
\bauthor{\bsnm{Fenton}, \binits{F.}},
\bauthor{\bsnm{Filippi}, \binits{S.}},
\bauthor{\bsnm{Veneziani}, \binits{A.}}:
\batitle{Experimental validation of a variational data assimilation procedure
  for estimating space-dependent cardiac conductivities}.
\bjtitle{Computer Methods in Applied Mechanics and Engineering}
\bvolume{358},
\bfpage{112615}
(\byear{2020})
\end{barticle}
\endbibitem

\bibitem{Irakoze2021Dist}
\begin{barticle}
\bauthor{\bsnm{Irakoze}, \binits{{\'E}.}},
\bauthor{\bsnm{Jacquemet}, \binits{V.}}:
\batitle{Multiparameter optimization of nonuniform passive diffusion properties
  for creating coarse-grained equivalent models of cardiac propagation}.
\bjtitle{Computers in Biology and Medicine}
\bvolume{138},
\bfpage{104863}
(\byear{2021}).
\doiurl{10.1016/j.compbiomed.2021.104863}
\end{barticle}
\endbibitem

\bibitem{chegini2021efficient}
\begin{barticle}
\bauthor{\bsnm{Chegini}, \binits{F.}},
\bauthor{\bsnm{Kopani{\v{c}}{\'a}kov{\'a}}, \binits{A.}},
\bauthor{\bsnm{Krause}, \binits{R.}},
\bauthor{\bsnm{Weiser}, \binits{M.}}:
\batitle{Efficient identification of scars using heterogeneous model
  hierarchies}.
\bjtitle{EP Europace}
\bvolume{23}(\bissue{Supplement\_1}),
\bfpage{113}--\blpage{122}
(\byear{2021})
\end{barticle}
\endbibitem

\bibitem{Arevalo2016}
\begin{barticle}
\bauthor{\bsnm{Arevalo}, \binits{H.J.}},
\bauthor{\bsnm{Vadakkumpadan}, \binits{F.}},
\bauthor{\bsnm{Guallar}, \binits{E.}},
\bauthor{\bsnm{Jebb}, \binits{A.}},
\bauthor{\bsnm{Malamas}, \binits{P.}},
\bauthor{\bsnm{Wu}, \binits{K.C.}},
\bauthor{\bsnm{Trayanova}, \binits{N.A.}}:
\batitle{{Arrhythmia risk stratification of patients after myocardial
  infarction using personalized heart models}}.
\bjtitle{Nature Communications}
\bvolume{7}(\bissue{May}),
\bfpage{11437}
(\byear{2016}).
\doiurl{10.1038/ncomms11437}
\end{barticle}
\endbibitem

\bibitem{sharp2019vector}
\begin{barticle}
\bauthor{\bsnm{Sharp}, \binits{N.}},
\bauthor{\bsnm{Soliman}, \binits{Y.}},
\bauthor{\bsnm{Crane}, \binits{K.}}:
\batitle{The vector heat method}.
\bjtitle{ACM Transactions on Graphics (TOG)}
\bvolume{38}(\bissue{3}),
\bfpage{1}--\blpage{19}
(\byear{2019})
\end{barticle}
\endbibitem

\bibitem{chambolle_introduction_2016}
\begin{barticle}
\bauthor{\bsnm{Chambolle}, \binits{A.}},
\bauthor{\bsnm{Pock}, \binits{T.}}:
\batitle{An introduction to continuous optimization for imaging}.
\bjtitle{Acta Numerica}
\bvolume{25},
\bfpage{161}--\blpage{319}
(\byear{2016}).
\doiurl{10.1017/S096249291600009X}.
Accessed 2018-10-05
\end{barticle}
\endbibitem

\bibitem{tensorflow}
\begin{bchapter}
\bauthor{\bsnm{Abadi}, \binits{M.}},
\bauthor{\bsnm{Barham}, \binits{P.}},
\bauthor{\bsnm{Chen}, \binits{J.}},
\bauthor{\bsnm{Chen}, \binits{Z.}},
\bauthor{\bsnm{Davis}, \binits{A.}},
\bauthor{\bsnm{Dean}, \binits{J.}},
\bauthor{\bsnm{Devin}, \binits{M.}},
\bauthor{\bsnm{Ghemawat}, \binits{S.}},
\bauthor{\bsnm{Irving}, \binits{G.}},
\bauthor{\bsnm{Isard}, \binits{M.}},
\bauthor{\bsnm{Kudlur}, \binits{M.}},
\bauthor{\bsnm{Levenberg}, \binits{J.}},
\bauthor{\bsnm{Monga}, \binits{R.}},
\bauthor{\bsnm{Moore}, \binits{S.}},
\bauthor{\bsnm{Murray}, \binits{D.G.}},
\bauthor{\bsnm{Steiner}, \binits{B.}},
\bauthor{\bsnm{Tucker}, \binits{P.}},
\bauthor{\bsnm{Vasudevan}, \binits{V.}},
\bauthor{\bsnm{Warden}, \binits{P.}},
\bauthor{\bsnm{Wicke}, \binits{M.}},
\bauthor{\bsnm{Yu}, \binits{Y.}},
\bauthor{\bsnm{Zheng}, \binits{X.}}:
\bctitle{Tensorflow: A system for large-scale machine learning}.
In: \bbtitle{12th USENIX Symposium on Operating Systems Design and
  Implementation (OSDI 16)},
pp. \bfpage{265}--\blpage{283}
(\byear{2016})
\end{bchapter}
\endbibitem

\bibitem{kingma_adam_2017}
\begin{botherref}
\oauthor{\bsnm{Kingma}, \binits{D.P.}},
\oauthor{\bsnm{Ba}, \binits{J.}}:
Adam: {A} {Method} for {Stochastic} {Optimization}.
arXiv:1412.6980 [cs]
(2017).
arXiv: 1412.6980.
Accessed 2020-07-29
\end{botherref}
\endbibitem

\bibitem{gharaviri2020epicardial}
\begin{barticle}
\bauthor{\bsnm{Gharaviri}, \binits{A.}},
\bauthor{\bsnm{Bidar}, \binits{E.}},
\bauthor{\bsnm{Potse}, \binits{M.}},
\bauthor{\bsnm{Zeemering}, \binits{S.}},
\bauthor{\bsnm{Verheule}, \binits{S.}},
\bauthor{\bsnm{Pezzuto}, \binits{S.}},
\bauthor{\bsnm{Krause}, \binits{R.}},
\bauthor{\bsnm{Maessen}, \binits{J.G.}},
\bauthor{\bsnm{Auricchio}, \binits{A.}},
\bauthor{\bsnm{Schotten}, \binits{U.}}:
\batitle{Epicardial fibrosis explains increased endo--epicardial dissociation
  and epicardial breakthroughs in human atrial fibrillation}.
\bjtitle{Frontiers in physiology}
\bvolume{11},
\bfpage{68}
(\byear{2020})
\end{barticle}
\endbibitem

\bibitem{fTetWild}
\begin{botherref}
\oauthor{\bsnm{Hu}, \binits{Y.}},
\oauthor{\bsnm{Schneider}, \binits{T.}},
\oauthor{\bsnm{Wang}, \binits{B.}},
\oauthor{\bsnm{Zorin}, \binits{D.}},
\oauthor{\bsnm{Panozzo}, \binits{D.}}:
Fast tetrahedral meshing in the wild.
ACM Trans.\ Graph.
\textbf{39}(4)
(2020).
\doiurl{10.1145/3386569.3392385}
\end{botherref}
\endbibitem

\bibitem{pezzuto2017evaluation}
\begin{barticle}
\bauthor{\bsnm{Pezzuto}, \binits{S.}},
\bauthor{\bsnm{Kal'avsk{\`y}}, \binits{P.}},
\bauthor{\bsnm{Potse}, \binits{M.}},
\bauthor{\bsnm{Prinzen}, \binits{F.W.}},
\bauthor{\bsnm{Auricchio}, \binits{A.}},
\bauthor{\bsnm{Krause}, \binits{R.}}:
\batitle{Evaluation of a rapid anisotropic model for {ECG} simulation}.
\bjtitle{Frontiers in physiology}
\bvolume{8},
\bfpage{265}
(\byear{2017})
\end{barticle}
\endbibitem

\bibitem{potse2018scalable}
\begin{barticle}
\bauthor{\bsnm{Potse}, \binits{M.}}:
\batitle{Scalable and accurate ecg simulation for reaction-diffusion models of
  the human heart}.
\bjtitle{Frontiers in physiology}
\bvolume{9},
\bfpage{370}
(\byear{2018})
\end{barticle}
\endbibitem

\bibitem{gillette2021framework}
\begin{barticle}
\bauthor{\bsnm{Gillette}, \binits{K.}},
\bauthor{\bsnm{Gsell}, \binits{M.A.}},
\bauthor{\bsnm{Prassl}, \binits{A.J.}},
\bauthor{\bsnm{Karabelas}, \binits{E.}},
\bauthor{\bsnm{Reiter}, \binits{U.}},
\bauthor{\bsnm{Reiter}, \binits{G.}},
\bauthor{\bsnm{Grandits}, \binits{T.}},
\bauthor{\bsnm{Payer}, \binits{C.}},
\bauthor{\bsnm{{\v{S}}tern}, \binits{D.}},
\bauthor{\bsnm{Urschler}, \binits{M.}}, \betal:
\batitle{A framework for the generation of digital twins of cardiac
  electrophysiology from clinical 12-leads ecgs}.
\bjtitle{Medical Image Analysis}
\bvolume{71},
\bfpage{102080}
(\byear{2021})
\end{barticle}
\endbibitem

\bibitem{lejeune2021exploring}
\begin{barticle}
\bauthor{\bsnm{Lejeune}, \binits{E.}},
\bauthor{\bsnm{Zhao}, \binits{B.}}:
\batitle{Exploring the potential of transfer learning for metamodels of
  heterogeneous material deformation}.
\bjtitle{Journal of the Mechanical Behavior of Biomedical Materials}
\bvolume{117},
\bfpage{104276}
(\byear{2021})
\end{barticle}
\endbibitem

\bibitem{wang2021eigenvector}
\begin{barticle}
\bauthor{\bsnm{Wang}, \binits{S.}},
\bauthor{\bsnm{Wang}, \binits{H.}},
\bauthor{\bsnm{Perdikaris}, \binits{P.}}:
\batitle{On the eigenvector bias of {F}ourier feature networks: From regression
  to solving multi-scale {PDEs} with physics-informed neural networks}.
\bjtitle{Computer Methods in Applied Mechanics and Engineering}
\bvolume{384},
\bfpage{113938}
(\byear{2021})
\end{barticle}
\endbibitem

\bibitem{Roney2021Fibers}
\begin{barticle}
\bauthor{\bsnm{Roney}, \binits{C.H.}},
\bauthor{\bsnm{Bendikas}, \binits{R.}},
\bauthor{\bsnm{Pashakhanloo}, \binits{F.}},
\bauthor{\bsnm{Corrado}, \binits{C.}},
\bauthor{\bsnm{Vigmond}, \binits{E.J.}},
\bauthor{\bsnm{McVeigh}, \binits{E.R.}},
\bauthor{\bsnm{Trayanova}, \binits{N.A.}},
\bauthor{\bsnm{Niederer}, \binits{S.A.}}:
\batitle{Constructing a human atrial fibre atlas}.
\bjtitle{Annals of Biomedical Engineering}
\bvolume{49}(\bissue{1}),
\bfpage{233}--\blpage{250}
(\byear{2021})
\end{barticle}
\endbibitem

\end{thebibliography}

\begin{appendices}

\section{Smooth basis}
\label{apx:smooth}

When defining a conduction velocity tensor through its eigencomponents, as was done in \Cref{sec:model}, any two non-coinciding (preferably orthonormal) vectors of the tangent space $\mathcal{S}$ may be a feasible choice.
Computationally, this 2D basis could be chosen locally, e.g. on a per-triangle basis.
However, as the chosen Huber-norm regularization in~\eqref{eq:loss} penalizes variation of the parameter vector $\vec{d}$, it is greatly beneficial to define a smooth basis.

To this end, we computed the first basis $\vec{v}_1$ of the tangent space using the vector heat method~\cite{sharp2019vector}.
The vector heat method quickly computes parallel transport of a chosen vector on the whole manifold by shortly diffusing a given initial vector field using the heat equation with the connection Laplacian $\Delta^{\nabla} = -\nabla_C^* \nabla_C$ for $\nabla_C$ being the Levi-Civita connection of the manifold.
The rescaled diffused vectors then closely approximate the parallel transport of the initial vector field (for numerical and qualitative comparisons, we refer to the original paper~\cite{sharp2019vector}).

The initial vector to be transported across the manifold is manually selected for each manifold, such that the discontinuities of the vector field are minimal.
Discontinuities might arise at the cut-locus, i.e.~regions equidistant from the given vector, as there are multiple possible parallel transports at such locations.
\begin{figure}[htb]
    \centering
    \includegraphics[width = \textwidth]{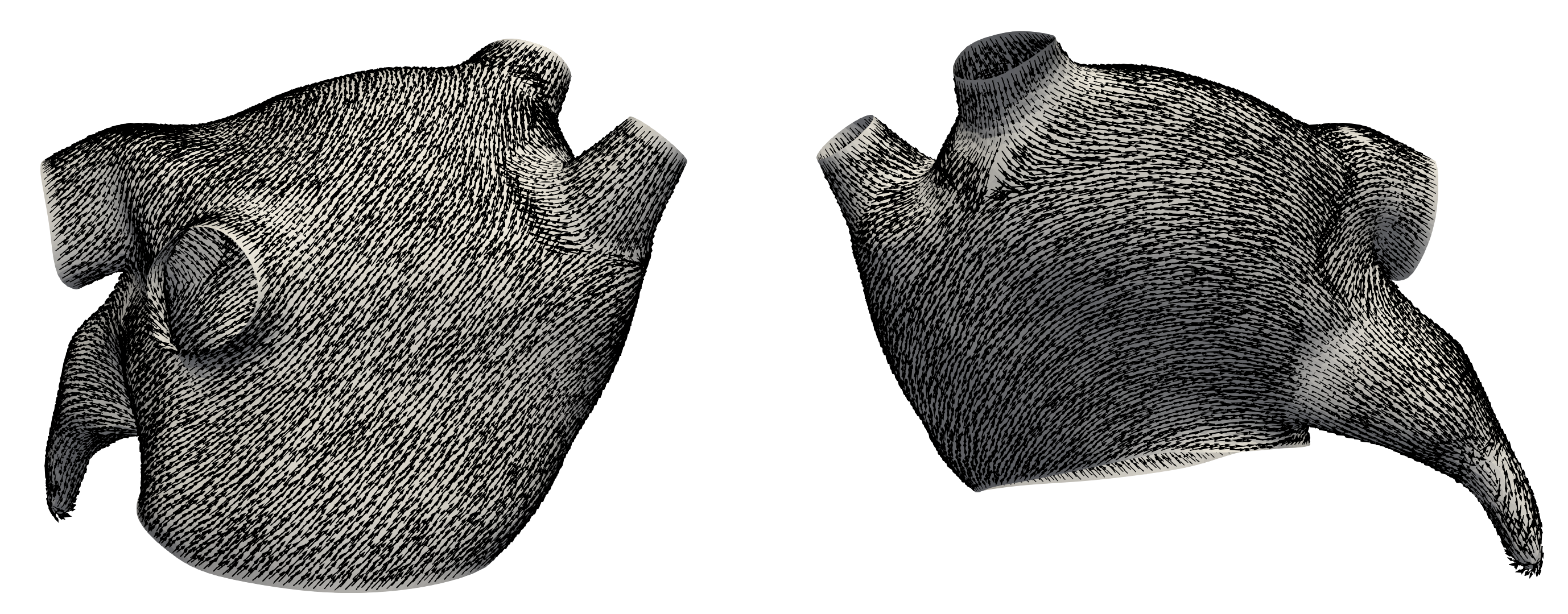}
    \caption{Result of transporting a single vector on the anterior wall of the atria across the entire manifold. 
    The resulting vector field is smooth except for parts of high curvature such as the left atrial appendage or pulmonary veins.}
    \label{fig:smoothbasis}
\end{figure}
\Cref{fig:smoothbasis} shows such a computed parallel transport from the anterior wall of the atria.
The resulting vector field is smooth in most parts, exhibiting a very low overall variation.
Such a basis was chosen manually for all meshes.

\section{DT-MR imaging errors}
\begin{table}[htb]
\centering
\resizebox{\columnwidth}{!}{
\begin{tabular}{c|c|ccccccc}
\toprule
\textbf{noise [ms]} & \textbf{\# of maps} & \textbf{case 1} & \textbf{case 2} & \textbf{case 3} & \textbf{case 4} & \textbf{case 5} & \textbf{case 6} & \textbf{case 7} \\
\midrule
0.0 & 1 & 31.4 (14.2 - 54.5) & 30.4 (13.4 - 55.4) & 26.2 (10.7 - 52.4) & 28.4 (11.9 - 55.3) & 26.9 (11.3 - 51.8) & 23.0 (9.7 - 46.6) & 21.5 (9.1 - 45.7)\\
0.0 & 3 & 22.5 (9.7 - 44.8) & 22.1 (9.6 - 44.6) & 19.7 (8.2 - 40.8) & 21.5 (9.2 - 42.2) & 21.6 (9.3 - 42.8) & 18.6 (7.7 - 38.5) & 17.4 (7.7 - 36.0)\\
0.0 & 5 & 21.6 (9.4 - 44.5) & 20.1 (8.6 - 41.9) & 17.5 (7.3 - 39.0) & 18.5 (8.0 - 38.4) & 19.8 (8.3 - 40.5) & 17.9 (7.7 - 38.1) & 17.0 (7.6 - 33.9)\\
\midrule
0.1 & 1 & 34.5 (15.6 - 60.4) & 26.5 (11.5 - 51.6) & 24.6 (11.1 - 48.9) & 30.7 (13.3 - 56.4) & 27.2 (11.2 - 52.6) & 28.0 (12.1 - 54.0) & 25.7 (11.3 - 49.8)\\
0.1 & 3 & 22.9 (10.1 - 45.5) & 24.1 (11.0 - 47.1) & 19.4 (8.2 - 40.2) & 20.9 (9.0 - 42.3) & 21.9 (9.6 - 43.8) & 19.7 (8.5 - 40.4) & 18.0 (7.9 - 36.4)\\
0.1 & 5 & 20.9 (9.1 - 43.7) & 22.1 (9.5 - 43.6) & 17.5 (7.3 - 39.0) & 19.7 (8.8 - 39.9) & 22.4 (10.1 - 45.0) & 19.8 (8.3 - 41.3) & 14.9 (6.5 - 31.2)\\
\midrule
1.0 & 1 & 45.6 (22.4 - 67.9) & 44.5 (22.0 - 67.0) & 37.2 (18.4 - 59.9) & 41.4 (19.4 - 65.6) & 43.6 (20.9 - 66.5) & 42.6 (20.8 - 65.1) & 42.5 (20.8 - 66.2)\\
1.0 & 3 & 38.6 (18.0 - 62.8) & 29.5 (12.9 - 54.3) & 30.7 (14.2 - 54.9) & 31.5 (14.4 - 55.2) & 30.5 (13.7 - 55.5) & 31.8 (14.3 - 55.2) & 30.6 (13.7 - 53.4)\\
1.0 & 5 & 33.0 (14.8 - 58.6) & 26.6 (11.9 - 50.5) & 24.6 (11.2 - 48.0) & 26.7 (11.2 - 50.4) & 25.2 (10.7 - 49.1) & 25.5 (11.5 - 47.8) & 26.0 (11.9 - 48.1)\\
\midrule
2.0 & 1 & 46.7 (25.3 - 68.0) & 43.6 (20.6 - 67.7) & 36.4 (18.1 - 59.8) & 43.2 (21.5 - 66.1) & 43.1 (19.3 - 65.9) & 42.8 (20.7 - 66.2) & 47.8 (24.6 - 69.7)\\
2.0 & 3 & 44.1 (22.0 - 67.2) & 41.8 (20.3 - 66.3) & 35.5 (18.0 - 59.7) & 40.5 (19.8 - 63.5) & 37.8 (18.3 - 61.0) & 37.4 (16.9 - 61.6) & 35.7 (16.8 - 58.9)\\
2.0 & 5 & 42.9 (21.9 - 64.8) & 41.1 (20.4 - 65.2) & 31.3 (13.6 - 55.8) & 35.8 (16.6 - 59.7) & 34.7 (15.8 - 59.5) & 33.7 (15.4 - 57.6) & 34.8 (16.1 - 59.7)\\
\bottomrule
\end{tabular}}

\caption{Fiber errors for the 7 fiber distributions obtained from DT-MR imaging. Results are presented for different levels of noise standard deviation and the number of maps used for training. Results are presented as median (25\% - 75\% percentile).}
\label{tab:atlaserrors}
\end{table}

\end{appendices}

\end{document}